\documentclass[%
 reprint,
superscriptaddress,
 amsmath,amssymb,
 aps,
floatfix,
]{revtex4-2}

\usepackage{graphicx}% Include figure files
\usepackage{dcolumn}% Align table columns on decimal point
\usepackage{bm}% bold math

\usepackage{amsmath}
\usepackage{lmodern}
\usepackage{siunitx}
\usepackage{booktabs}
\usepackage{etoolbox}
\usepackage{tabularx}
\usepackage{placeins}
\usepackage{booktabs}
\usepackage{multirow}
\usepackage{threeparttable}
\usepackage{physics}
\usepackage{url}
\usepackage{float}

\usepackage[hidelinks]{hyperref}

\newcommand{\RomanNumeralCaps}[1]
    {\MakeUppercase{\romannumeral #1}}

\newcommand{\mycomment}[1]{}

\begin{document}

\preprint{APS/123-QED}

\title{Machine-Learned Bond-Order Potential for Exploring the Configuration Space of Carbon}% Force line breaks with \\

\author{Ikuma Kohata}
\altaffiliation{Contact author: kohata@photon.t.u-tokyo.ac.jp}
\affiliation{Department of Mechanical Engineering, The University of Tokyo, 7-3-1 Hongo, Bunkyo-ku, Tokyo 113-8656, Japan}

\author{Kaoru Hisama}
\affiliation{Research Initiative for Supra-Materials, Shinshu University, 4-17-1 Wakasato, Nagano, Nagano 380-8553, Japan}

\author{Keigo Otsuka}
\affiliation{Department of Mechanical Engineering, The University of Tokyo, 7-3-1 Hongo, Bunkyo-ku, Tokyo 113-8656, Japan}

\author{Shigeo Maruyama}
\affiliation{Department of Mechanical Engineering, The University of Tokyo, 7-3-1 Hongo, Bunkyo-ku, Tokyo 113-8656, Japan}

\date{\today}% It is always \today, today,

\begin{abstract}
Construction of transferable machine-learning interatomic potentials with a minimal number of parameters is important for their general applicability.
Here, we present a machine-learning interatomic potential with the functional form of the bond-order potential for comprehensive exploration over the configuration space of carbon. The physics-based design of this potential enables robust and accurate description over a wide range of the potential energy surface with a small number of parameters. 
We demonstrate the versatility of this potential through validations across various tasks, including phonon dispersion calculations, global structure searches for clusters, phase diagram calculations, and enthalpy-volume mappings of local minima structures. 
We expect that this potential can contribute to the discovery of novel carbon materials.
\end{abstract}

\maketitle

%\tableofcontents

%\section{\label{sec:level1}First-level heading:\protect\\ The line
%break was forced \lowercase{via} \textbackslash\textbackslash}

\section{Introduction}
Because of its variety of bonding states, carbon can form various allotropes with very unique physical properties such as diamond, fullerene \cite{Kroto1985}, graphene \cite{doi:10.1126/science.1102896}, and carbon nanotubes \cite{iijima1991helical,1993Natur.363..603I}. 
These carbon allotropes are actively studied, and at the same time, the prediction of novel carbon allotropes is of great interest. 
In the past, computational studies based on density functional theory (DFT) have predicted new hypothetical carbon allotropes such as Z-carbon \cite{PhysRevLett.108.065501}, M-carbon \cite{PhysRevLett.102.175506}, bct-C4 \cite{PhysRevLett.104.125504}, and S-carbon \cite{HE20121560}.
However, the computational cost of DFT is a bottleneck in the further comprehensive search for a large number of structures without restrictions on the system size.

Empirical interatomic potentials are computationally inexpensive methods for calculating the physical properties of materials without explicitly simulating electronic structures. Among these, bond-order potentials are a class of empirical potentials designed to model the behavior of covalently bonded elements. The concept of bond-order potential was originally proposed by Abell \cite{PhysRevB.31.6184} and specifically formulated by Tersoff \cite{PhysRevLett.56.632,PhysRevB.38.9902,PhysRevB.37.6991}. In bond-order potentials, the potential energy of the system $E$ is given by
\begin{align}
    E &= \frac{1}{2}\sum_{i}\sum_{j\neq i}V_{ij}, \\
    V_{ij} &= a_{ij}V_{R}(r_{ij}) - b_{ij}V_{A}(r_{ij}).
\end{align}
Here $i$, $j$ index the atoms in the system, $r_{ij}$ represents the distance between atom $i$ and atom $j$, $V_{ij}$ is the energy assigned to bond $i$-$j$, and $V_{R}$ and $V_{A}$ are the repulsive and attractive terms of a Morse-type pair potential. The parameters $a_{ij}$ and $b_{ij}$ scale the attractive and repulsive terms and are 
computed based on the local atomic environments.
This bond-order potential, called Tersoff potential, has been parameterized for silicon \cite{PhysRevLett.56.632,PhysRevB.38.9902,PhysRevB.37.6991}, carbon \cite{PhysRevLett.61.2879}, and multicomponent systems of group-\RomanNumeralCaps{4} elements \cite{PhysRevB.39.5566,PhysRevB.41.3248.2,PhysRevLett.64.1757}, and many variations have been proposed over the decades, such as the Reactive Empirical Bond-Order (REBO) potential \cite{PhysRevB.42.9458,PhysRevB.46.1948.2,DonaldWBrenner_2002}, the Adaptive Intermolecular Reactive Empirical Bond-Order (AIREBO) potential \cite{10.1063/1.481208}, the long-range carbon bond-order potential (LCBOP) \cite {PhysRevB.68.024107,PhysRevB.72.214102,PhysRevB.73.229901,PhysRevB.72.214103}, and several modified functional forms of the Tersoff potential \cite{KUMAGAI2007457,PhysRevB.78.161402,PhysRevB.87.205410,PhysRevB.97.125411,doi:10.1021/acs.jpcc.7b12687}.
Although these bond-order potentials offer simple description and fast computation of potential energy, their limited number of parameters constrains their flexibility, thereby limiting the range of structures and properties they can accurately describe and their applicability to material discovery.

In recent years, the emergence of machine learning interatomic potentials (MLIPs) has made significant progress in computational materials science. MLIPs offer a more flexible representation of the potential energy surface, enabling the description of diverse properties and structures with a single parameter set. 
Carbon has been one of the main targets of MLIPs for years. In early studies, MLIPs for carbon were developed by Khaliullin \textit{et al.} \cite{PhysRevB.81.100103} based on the neural network representation of the potential energy surface proposed by Behler and Parrinello \cite{PhysRevLett.98.146401} and by Bart\'ok \textit{et al.} \cite{PhysRevLett.104.136403} based on the Gaussian Approximation Potential (GAP), demonstrating the effectiveness of MLIPs in describing the complex covalent bonding systems.
Starting from these studies, many MLIPs for carbon were developed using various models, including Behler-Parrinello type neural network potentials (NNPs) \cite{Shaidu2021,Cheng2023}, SchNet \cite{HEDMAN2021100027}, DeePMD \cite{WANG20221}, GAP \cite{PhysRevB.95.094203,10.1063/5.0005084,10.1063/5.0091698,doi:10.1021/acs.chemmater.1c03279}, Atomic Cluster Expansion (ACE) \cite{doi:10.1021/acs.jctc.2c01149}, and the spectral neighbor analysis potential (SNAP) \cite{PhysRevB.106.L180101}. 
Despite these advances, achieving comprehensive coverage of the atomic configurations and phase diagram of carbon over a wide range of pressures with a single set of the limited number of parameters remains difficult due to the vast configuration space that must be captured.

One major obstacle in describing the entire potential energy surface is the poor transferability of MLIPs; they are often inaccurate outside of the configuration space covered by the training dataset. One way to overcome this and build MLIPs with good transferability is to reduce the model complexity. This is justified by the principle of Occam's razor \cite{Lang:alg}, which states that simpler explanations or models should be preferred to more complex ones if both explain the data equally well. A key feature of potential energy surfaces is that two-body interactions exhibit much simpler behavior than many-body interactions, even though interatomic distance is the dominant factor determining the potential energy. In accordance with Occam’s razor, it is therefore reasonable to model two-body and many-body interactions separately, using models of appropriate complexity for each, as is done in bond-order potentials.

In this study, motivated by these issues, we present an MLIP model that combines a machine-learned bond-order with a classical two-body energy function. This model, named Machine-Learning Bond-Order Potential (MLBOP), provides a robust and accurate description of the potential energy surface with a small number of parameters.
The idea of combining the traditional potential energy function with machine-learning was previously discussed by Pun \textit{et al.}, who developed the Physically Informed Neural Network (PINN) potential \cite{Pun2019}, an MLIP model with a combination of the bond-order potential and the machine-learned on-the-fly correction of potential parameters depending on the local environments. The PINN model was used to develop general-purpose interatomic potentials for aluminum \cite{Pun2019,PhysRevMaterials.4.113807} and tantalum \cite{LIN2022111180}, and demonstrated excellent robustness in extrapolation domains. 
The MLBOP model proposed here adopts the philosophy of the PINN model, incorporating some modifications to enhance functionality and interpretability.
In the PINN model, potential parameters are obtained by summing preoptimized global parameters with machine-learned local parameters, making all potential parameters dependent on the local environment. In contrast, the MLBOP model predetermines parameters in the repulsive and attractive two-body terms based on chemical species, training only the bond-order against local environments. This approach strictly adheres to Abell's binding energy representation, while improving interpretability and simplifying both the model structure and the training process.
Furthermore, the MLBOP model incorporates a dispersion energy term expressed as an analytical function to enhance its applicability to systems where van der Waals interactions play important roles, such as layered materials. 

Here we use this model to develop a general-purpose MLIP for carbon that correctly captures the continuity and smoothness of the potential energy surface over a wide range compared to other reported MLIPs.
In section \ref{sec:dev}, we show the formalism and parameter optimization protocol of MLBOP.
In section \ref{sec:Robust}, we test the robustness of MLBOP by training it on several training datasets and checking the potential energy curves calculated with MLBOP.
In section \ref{sec:training}, we describe the dataset construction and training results of the general-purpose MLBOP for carbon.
In section \ref{sec:verification}, we evaluate the general-purpose MLBOP on basic physical properties of several systems, including crystal, defect, surface, liquid, and amorphous structures, and then assess the applicability of MLBOP to more advanced tasks, such as global structure searches for clusters, phase diagram calculations, and enthalpy-volume mapping of the local minima structures. 
Finally, in section \ref{sec:conclusion}, we summarize this work.

\section{Method}
\label{sec:dev}
\subsection{Formulation of MLBOP}
The formulation of MLBOP presented in this study basically follows Tersoff’s expression of bond-order \cite{PhysRevLett.56.632,PhysRevB.38.9902,PhysRevB.37.6991}, with the nonlinear functions replaced by multilayer perceptrons (MLPs) while preserving the original structural framework.
In the MLBOP model, the potential energy of the system $E_\mathrm{tot}$ is calculated as a sum of the bond energy $E_\mathrm{bond}$ and the dispersion energy $E_\mathrm{disp}$:
\begin{equation} 
E_\mathrm{tot} = E_\mathrm{bond} + E_\mathrm{disp}.
\label{eq:dftd3}
\end{equation}

\begin{figure}[htb!]
\centering
\includegraphics[clip,scale=0.54]{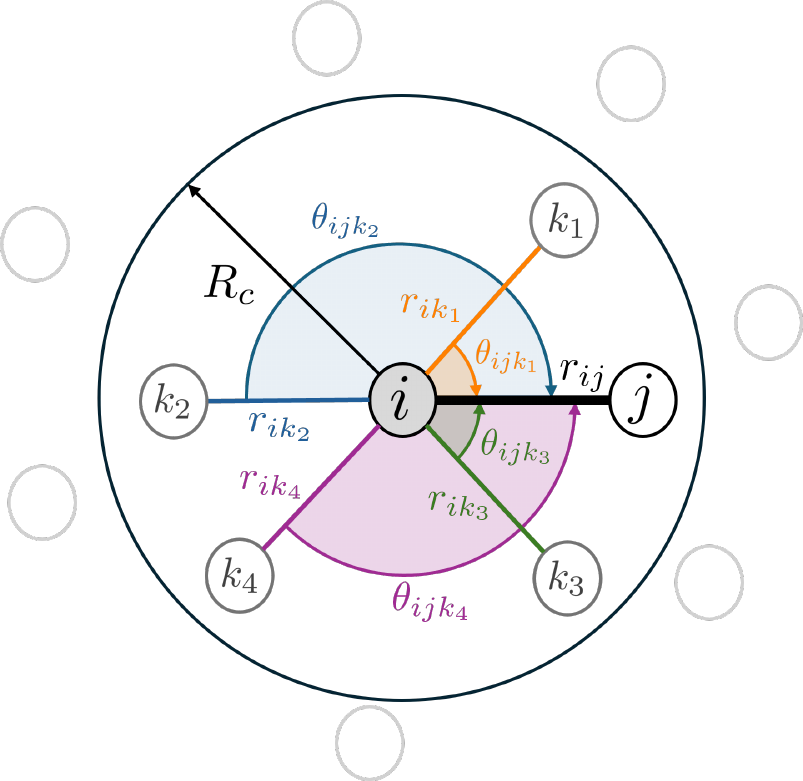}
\caption{Schematic diagram of the local environment of bond $i$-$j$ surrounded by atoms $k_{1}$, $k_{2}$, $k_{3}$, and $k_{4}$ within the cutoff distance $R_{c}$.}
\label{fig:threebody}
\end{figure}

The form of $E_\mathrm{bond}$ can be written as
\begin{align} 
E_\mathrm{bond} &= \frac{1}{2}\sum_{i}\sum_{j \neq i}[a_{ij}V_{R}(r_{ij}) - b_{ij}V_{A}(r_{ij})] ,
\end{align} 
where $r_{ij}$, $V_{R}(r_{ij})$, and $V_{A}(r_{ij})$ are the interatomic distance between atom $i$ and atom $j$, the repulsive energy shown in Eq.\ref{eq:repulsive}, and the attractive energy shown in Eq.\ref{eq:attractive}, respectively. The two-body terms $V_{R}(r_{ij})$ and $V_{A}(r_{ij})$ depend only on the bond length, while the values $a_{ij}$ and $b_{ij}$ depend on the local environment around bond $i$-$j$.
The forms of the repulsive energy $V_{R}$ and the attractive energy $V_{A}$ can be written as
\begin{equation} 
V_{R}(r_{ij}) = f_{c1}(r_{ij}) \left ( 1+\frac{Q}{r_{ij}} \right) \sum_{n=1}^{3} A_{n}\exp(-\alpha_{n}r_{ij}),
\label{eq:repulsive}
\end{equation}
and
\begin{equation} 
V_{A}(r_{ij}) = f_{c1}(r_{ij}) \sum_{n=1}^{3} B_{n}\exp(-\beta_{n}r_{ij}),
\label{eq:attractive}
\end{equation}
where $Q$, $A_{n}$, $\alpha_{n}$, $B_{n}$, and $\beta_{n}$ are the trainable parameters, and $f_{c1}$ is the cutoff function that truncates the interaction between atoms $i$ and $j$ at the cutoff distance $R_{c}$, as defined in Eq. \ref{eq:cutoff}.
The parameters $a_{ij}$ and $b_{ij}$ are computed using the sequence of MLPs from the local environment of bond $i$-$j$, defined by the spatial configuration of neighboring atoms within a sphere of radius $R_c$ centered on atom $i$, as illustrated in Figure~\ref{fig:threebody}.

\begin{figure}[htb!]
\centering
\includegraphics[clip,scale=0.40]{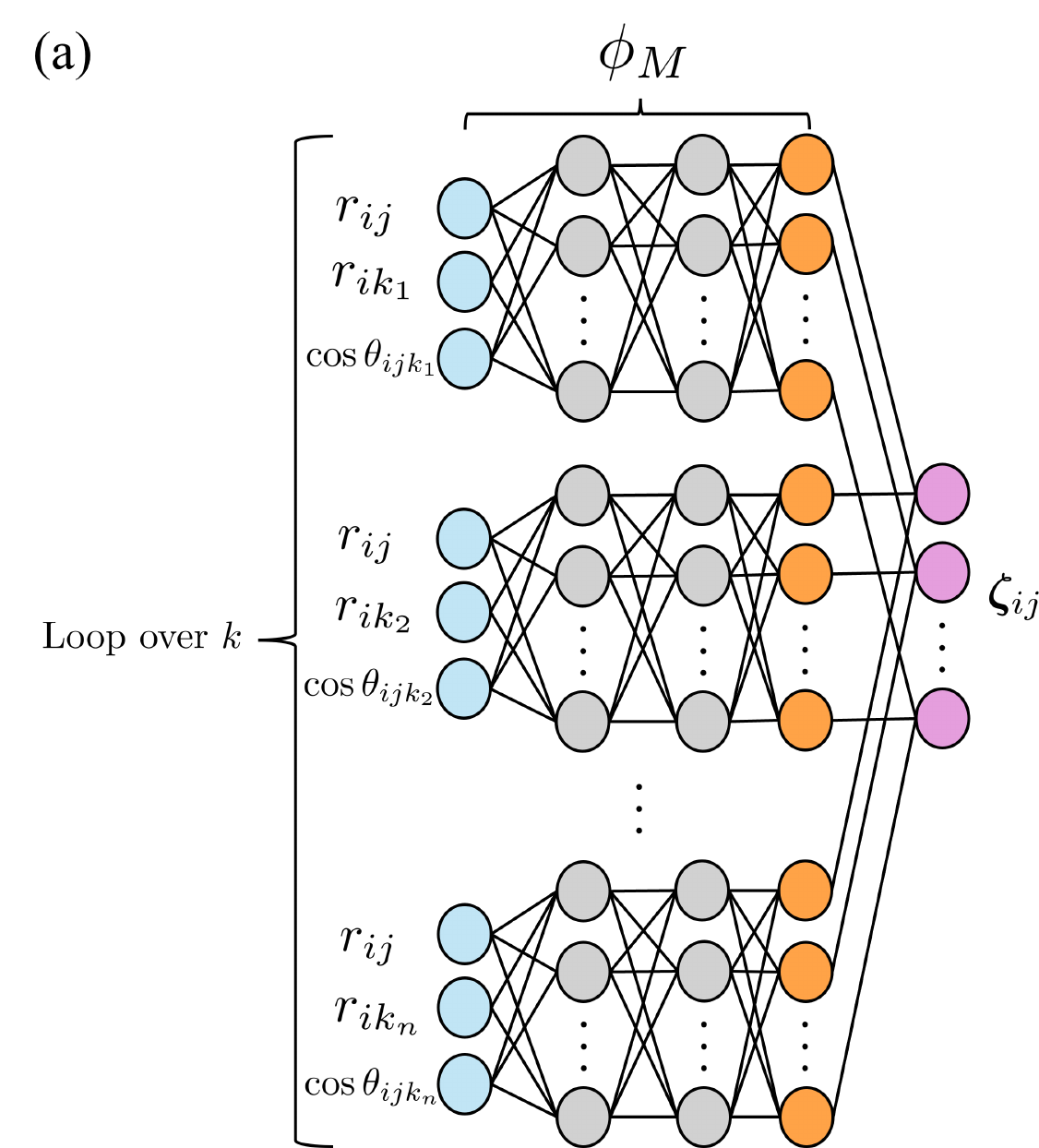}
\includegraphics[clip,scale=0.40]{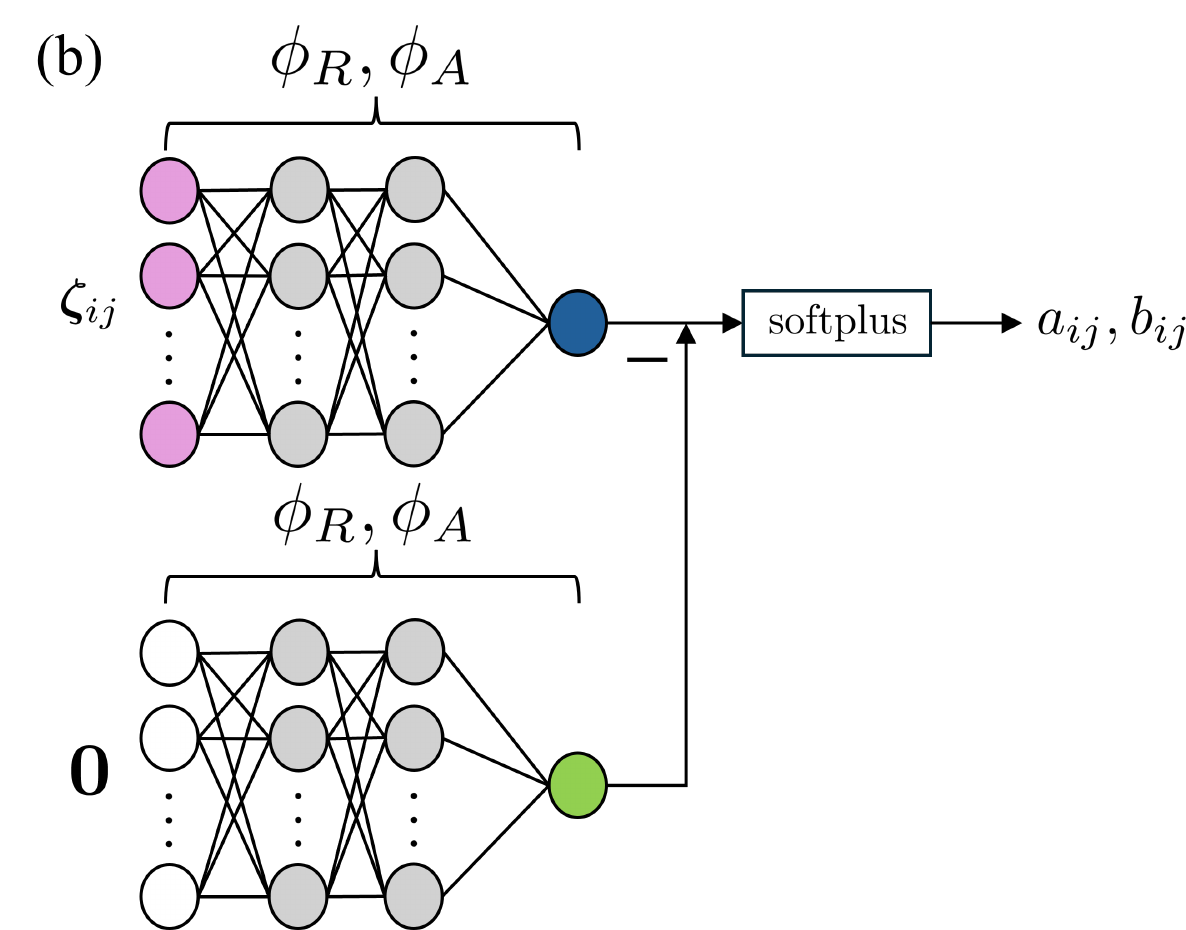}
\caption{Schematic structure of the MLBOP model. (a) Construction of the bond-embedded feature vector $\bm{\zeta}_{ij}$ in Eq. \ref{eq:zeta}. (b) Computation process of $a_{ij}$ and $b_{ij}$ in Eq. \ref{eq:aij} and Eq. \ref{eq:bij}.}
\label{fig:sketch}
\end{figure}

The model architecture used to compute $a_{ij}$ and $b_{ij}$ is illustrated in Figure~\ref{fig:sketch}. The bond-embedded descriptor $\bm{\zeta}_{ij}$, which captures the local environment around bond $i$–$j$, is defined as
\begin{equation}
\bm{\zeta}_{ij} = \sum_{k \neq i,j} f_{c2}(r_{ik})  \phi_M(r_{ij}, r_{ik}, \cos\theta_{ijk}),
\label{eq:zeta}
\end{equation}
where $\theta_{ijk}$ is the angle between bonds $i$–$j$ and $i$–$k$, $f_{c2}$ is a cutoff function, and $\phi_M$ is an MLP that maps the three-body configuration $(r_{ij}, r_{ik}, \cos\theta_{ijk})$ to a higher-dimensional vector. The sum is taken over all neighboring atoms $k$ of atom $i$, with $k = j$ excluded to remove self-interactions.

Using this descriptor, the parameters $a_{ij}$ and $b_{ij}$ are computed as
\begin{align}
a_{ij} &= \sigma\big( \phi_R(\bm{\zeta}_{ij}) - \phi_R(\bm{0}) \big), \label{eq:aij} \\
b_{ij} &= \sigma\big( \phi_A(\bm{\zeta}_{ij}) - \phi_A(\bm{0}) \big), \label{eq:bij}
\end{align}
where $\phi_R$ and $\phi_A$ are MLPs that map $\bm{\zeta}_{ij}$ to scalar values, and $\bm{0}$ is a zero vector of the same dimension as $\bm{\zeta}_{ij}$. The softplus activation function $\sigma$ is defined as
\begin{equation}
\sigma(x) = \frac{1}{\beta} \ln(1 + \exp(\beta x)),
\label{eq:sftp}
\end{equation}
with $\beta = \ln(2)$. To obtain $a_{ij}$ and $b_{ij}$, $\phi_{R}(\bm{\zeta}_{ij})$ and $\phi_{A}(\bm{\zeta}_{ij})$ are shifted by $\phi_{R}(\bm{0})$ and $\phi_{A}(\bm{0})$, and then transformed using the softplus function, ensuring that $a_{ij}$ and $b_{ij}$ remain positive and satisfy $a_{ij}=1$ and $b_{ij}=1$ at $\bm{\zeta}_{ij}=\bm{0}$. This guarantees that the potential energy of an isolated dimer is fixed to the two-body energy, $V_{R}(r)-V_{A}(r)$.
The cutoff function $f_{c1}$ and $f_{c2}$ are calculated as
\begin{equation} 
f_{c1,2}(r) = \left [  \dfrac{\tanh \{ B_{c1,2}(1-r/R_{c}) \} }{\tanh(B_{c1,2})} \right ]^{3} \; (r<R_{c}),
\label{eq:cutoff}
\end{equation}
where $R_{c}$ is the cutoff distance and $B_{c1,2}$ are the trainable parameters, which have different values for $f_{c1}$ and $f_{c2}$. The cutoff functions and their first and second derivative vanish at the cutoff distance $R_{c}$. 
In the following sections, the architectures of the MLPs $\phi_{M}$, $\phi_{R}$, and $\phi_{A}$ are represented as hyphenated sequences (e.g., 3-5-10-20), where each number indicates the number of units in each layer, listed from the input layer to the output layer.

The dispersion energy term $E_\mathrm{disp}$ is calculated based on the Grimme's DFT-D3 method \cite{10.1063/1.3382344} with the Becke-Johnson dumping function \cite{https://doi.org/10.1002/jcc.21759}, and can be written as
\begin{equation} 
E_\mathrm{disp} = -\frac{1}{2}\sum_{i}\sum_{j \neq i}\left( \frac{s_{6}C_{6ij}}{r_{ij}^6+f_{d}(R_{0})^6}+\frac{s_{8}C_{8ij}}{r_{ij}^8+f_{d}(R_{0})^8} \right)
\end{equation}
with
\begin{equation} 
f_{d}(R_{0})=a_{1}R_{0}+a_{2}.
\end{equation}
Here $s_{6}$, $s_{8}$, $a_{1}$, and $a_{2}$ are preoptimized parameters depending on the exchange-correlation functional, $R_{0}$ is a preoptimized parameter depending on the elements of atom $i$ and atom $j$. The values $C_{6ij}$ and $C_{8ij}$ are the dispersion coefficients calculated from the coordination numbers around atom $i$ and atom $j$.

\subsection{Protocol for parameter optimization}

\label{subsec:Param}

In order to train MLBOP, we optimize an absolute combined loss function
\begin{equation}
\begin{split}
L(\bm{w}) = \rho \sum_{p=1}^{N}|(E_{p}-g n_{p})-(\hat{E}_{p}(\bm{w})-\hat{g}(\bm{w}) n_{p})| \\
+\sum_{p=1}^{N}\left(\frac{1}{3n_{p}}\sum_{i=1}^{n_{p}}||\bm{F}_{p,i}-\hat{\bm{F}}_{p,i}(\bm{w})||\right)
\end{split}
\label{eq:lossfunc}
\end{equation} 
where $\bm{w}$ represents the trainable parameters, $N$ is the mini-batch size, $p$ indexes each structure, and $n_p$ denotes the number of atoms in structure $p$. $\rho$ is the energy loss weight. The DFT energy of structure $p$ and the atomic forces on atom $i$ in structure $p$ are denoted by $E_p$ and $\bm{F}_{p,i}$, respectively, while the predicted energy and forces are represented by $\hat{E}_p$ and $\hat{\bm{F}}_{p,i}$. The terms $g$ and $\hat{g}$ correspond to the per-atom energy of a ground-state structure calculated with DFT and MLBOP, respectively. %These terms are added to the loss function to compensate the energy difference coming from the zero-point energy.

The two-body and three-body parameters in $E_\mathrm{bond}$ are trained in separate stages. 
The two-body parameters in $V_{R}$ and $V_{R}$ are trained on the potential energy vs bond length curve of an isolated dimer.
The three-body parameters in $a_{ij}$ and $b_{ij}$ are then trained on the training dataset, while the two-body parameters are fixed.
The weights in the MLPs are stochastically initialized using Kaiming's method \cite{he2015delvingdeeprectifierssurpassing}, and the biases are initialized to zero.
The parameter optimization is performed by the mini-batch stochastic gradient descent using the ADAM optimizer \cite{kingma2017adam}. The energy loss weight $\rho$ is set to 0.1.
The mini-batch size is set to 4. The initial learning rate is set to $1.0\times10^{-3}$ and reduced to 10\% each time a loss shows no improvement for 10 epochs. The parameter optimization is continued until the learning rate reaches $1.0\times10^{-7}$. 

\section{Model Robustness on Potential energy surfaces\protect}
\label{sec:Robust}

\begin{figure}[htb!]
\centering
\includegraphics[clip,scale=0.37]{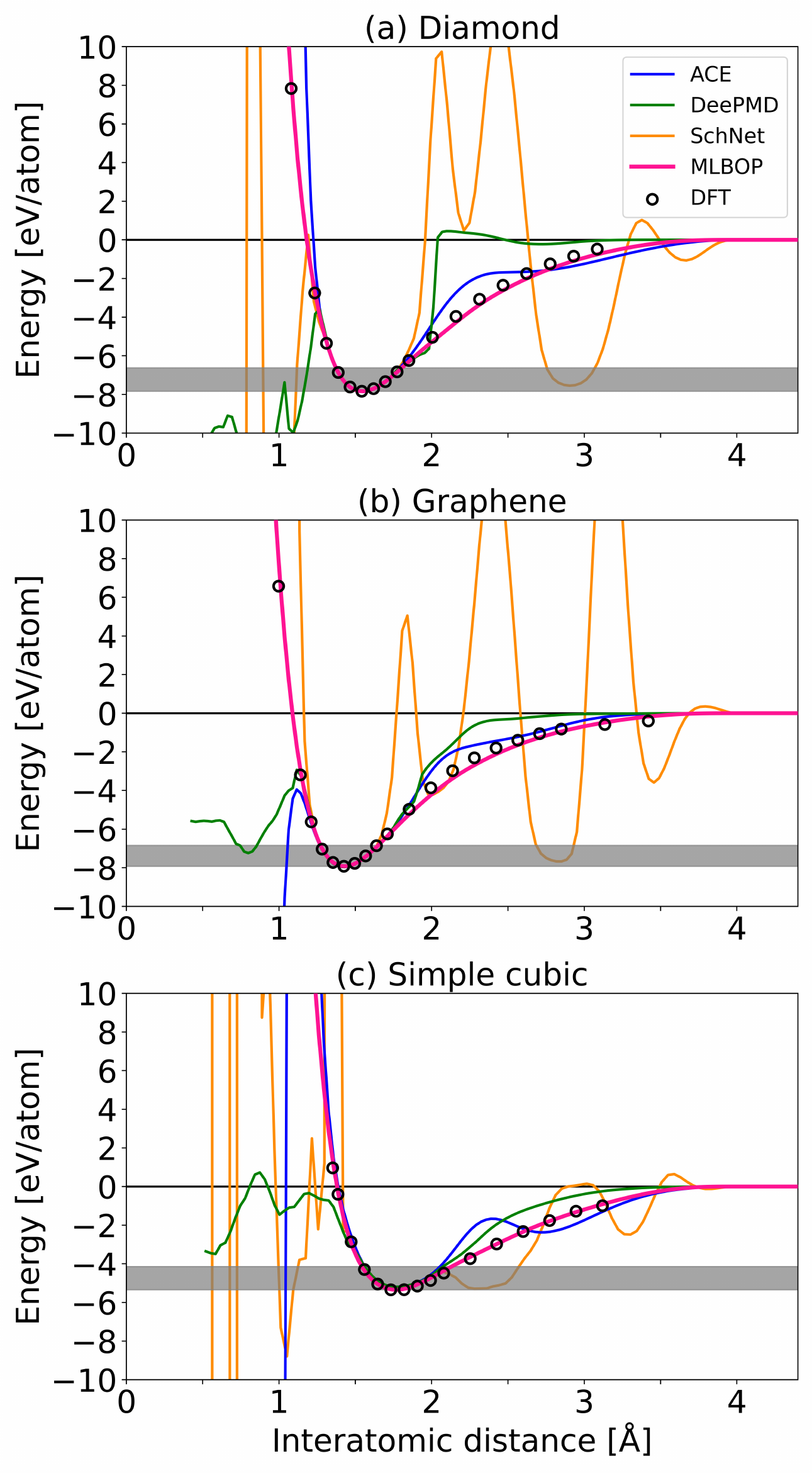}
\caption{Cohesive energy vs nearest-neighbor interatomic distance curves for (a) diamond, (b) graphene, and (c) simple cubic, calculated with MLBOP, DeePMD, ACE, and SchNet trained on the toy datasets, and DFT (PBE). The gray regions indicate the per-atom energy range covered by the training datasets.}
\label{fig:transferability}
\end{figure}

To test the model robustness in describing potential energy surfaces, we trained MLBOP on three toy datasets of carbon, each containing 400 diamond, graphene, and simple cubic structures with lattice constants ranging from 90\% to 110\% of their equilibrium values and with random atomic displacements up to 0.2~\r{A}. Each structure in the diamond, graphene, and simple cubic datasets has 64, 50, and 64 atoms, respectively.
An isolated atom was also added to each dataset to enable the models to learn the cohesive energy.
To obtain the reference DFT energies, we used the the generalized gradient approximation with the Perdew-Burke-Ernzerhof (PBE) functional \cite{PhysRevLett.77.3865,PhysRevLett.78.1396} implemented in Vienna ab initio simulation package (VASP) \cite{PhysRevB.47.558,Kresse_1994,PhysRevB.54.11169,PhysRevB.59.1758}.
For comparison, we also trained DeePMD \cite{PhysRevLett.120.143001}, SchNet~\cite{10.1063/1.5019779}, and ACE \cite{PhysRevB.99.014104} on the same datasets.
In MLBOP, the vector $\bm{\zeta}_{ij}$ is set to a dimensionality of 20. The networks $\phi_{R}$ and $\phi_{A}$ each consist of two hidden layers with 20 neurons, resulting in an architecture of 20-20-20-1. The network $\phi_{M}$ includes two hidden layers with 5 and 10 neurons, with its structure denoted as 3-5-10-20.
For all trained models, the cutoff distance was set to 4.0~\r{A}. ACE was regularized with an L1 parameter of $10^{-5}$.
Figure \ref{fig:transferability} shows the cohesive energy vs nearest-neighbor distance curves of graphene, diamond, and simple cubic structure calculated with DFT, DeePMD, SchNet, ACE, and MLBOP trained on each dataset. 
Outside the energy range covered by the training datasets, DeePMD, SchNet, and ACE show a discrepancy from the DFT. On the other hand, MLBOP gives a good fit to the DFT while preserving the shape of the potential energy curve derived from the Morse-type two-body function in both the interpolation and extrapolation domains.

\begin{figure*}[htb!]
\includegraphics[clip,scale=0.16]{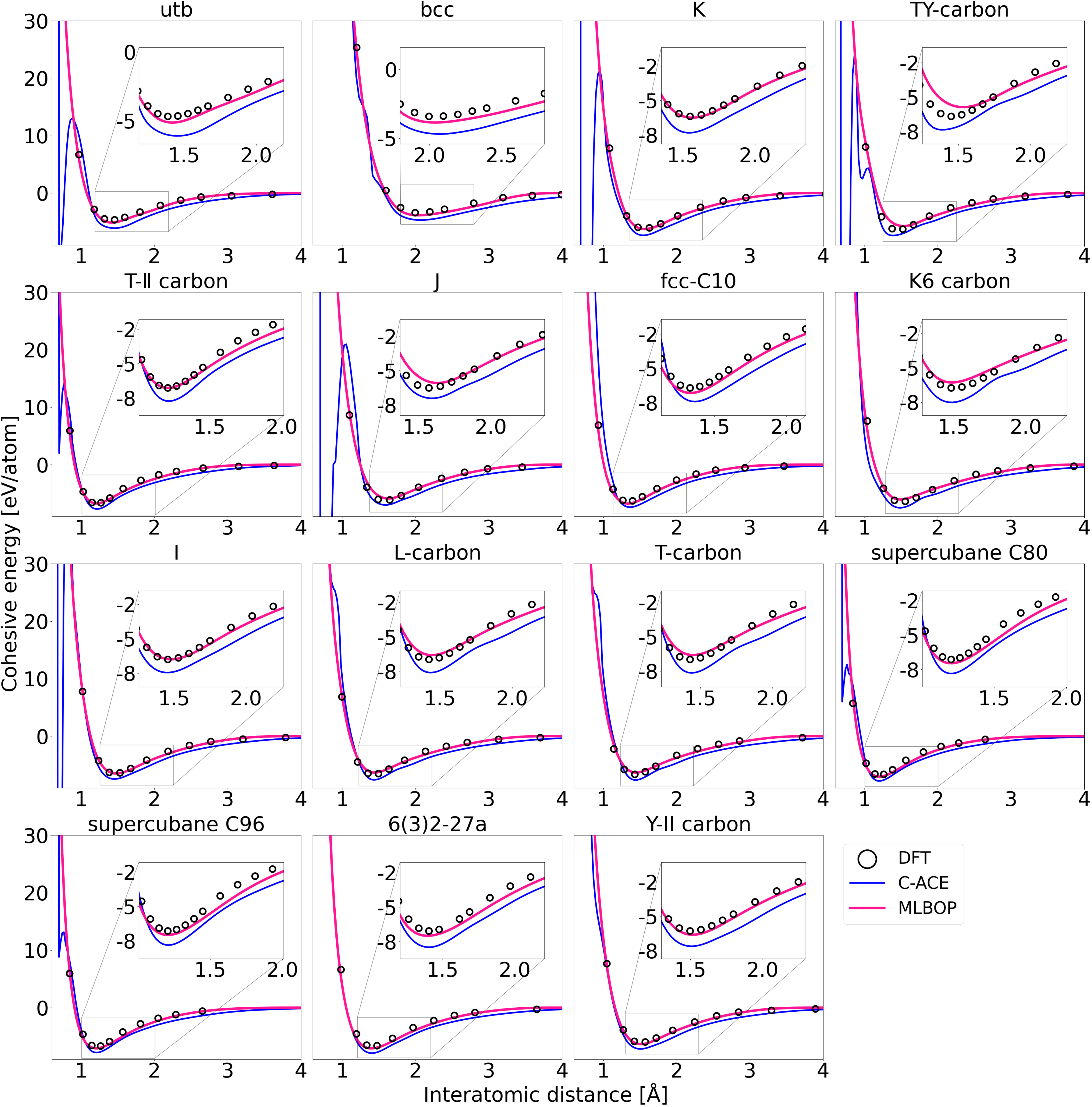}
\caption{Cohesive energy vs nearest-neighbor interatomic distance curves for the unstable crystals in the Samara Carbon Allotrope Database (SACADA) \cite{https://doi.org/10.1002/anie.201600655}, calculated with DFT (PBE), C-ACE, and MLBOP trained on the C-ACE dataset. The dispersion energy terms in MLBOP and C-ACE were omitted for comparison with the PBE energies. The boxes on the top right in each panel are magnified views of the low-energy regions.}
\label{fig:sacadarobust}
\end{figure*}

For further testing, we trained MLBOP on the dataset used to parametrize ACE for carbon (C-ACE) \cite{doi:10.1021/acs.jctc.2c01149}, which includes 17293 structures with various local environments. The root mean square errors (RMSEs) of MLBOP were 136 meV/atom for energy and 0.655 eV/\r{A} for force, comparable to those of C-ACE, which were 166 meV/atom and 0.689 eV/\r{A}, respectively, as listed in Table \ref{table:ace}. Figure \ref{fig:sacadarobust} shows the cohesive energy vs the nearest neighbor interatomic distance curves calculated with DFT, C-ACE, and MLBOP for the first to fifteenth most unstable crystals listed in the Samara Carbon Allotrope Database (SACADA) \cite{https://doi.org/10.1002/anie.201600655} as of September 2, 2024: utb (\#32 in SACADA), bcc (\#8), K (\#51), TY-carbon (\#34), T-\RomanNumeralCaps{2} carbon (\#35), J (\#117), fcc-C10 (\#152), K6 carbon (\#12), I (\#135), L-carbon (\#128), T-carbon (\#33), supercubane C80 (\#42), supercubane C96 (\#43), 6(3)2-27a (\#106), and Y-\RomanNumeralCaps{2} carbon (\#4). 
Compared to C-ACE, MLBOP reproduced the potential energy curve with excellent smoothness over the extrapolation domain for all tested structures while maintaining interpolation accuracy.

\begin{table}[htb!]
\caption{\label{table:ace}
Model comparison of RMSEs for energies $E_\mathrm{ace}$ and forces $F_\mathrm{ace}$ on the C-ACE datasets.}
\begin{ruledtabular}
\begin{tabular}{lcc}
Model &  $E_\mathrm{ace}$ (meV/atom) & $F_\mathrm{ace}$ (eV/\r{A}) \\ 
\midrule
MLBOP & 136 & 0.655 \\
ACE & 166 & 0.689 \\
\end{tabular}
\end{ruledtabular}
\end{table}

\section{Training a general-purpose MLBOP for carbon}
\label{sec:training}
\subsection{Building a training dataset}

In order to develop an interatomic potential for general purposes, the training dataset structures should cover as wide a range of the potential energy surface as possible. 
To achieve this, we collected structures with several categories: crystal, cluster, disordered, and random structure.
The crystal structures were generated by giving random atomic displacements and cell deformations to common crystal structures: graphene, graphite, nanotubes, diamond, simple cubic, body-centered cubic (bcc), face-centered cubic (fcc), and bc8. 
The crystal dataset structures also include defective structures generated by randomly removing atoms or rotating bonds, and surface structures with various surface orientations.
The cluster dataset structures were generated by randomly displacing atoms in linear chains and cyclic molecules with 3-12 atoms, and fullerenes with 20-80 atoms.
The disordered and random dataset structures were generated by iterative and somewhat \textit{ad hoc} expansion of the database using molecular dynamics (MD) simulation and random structure searching (RSS) \cite{Pickard_2011} performed with classical interatomic potentials and MLBOP under development.
The disordered structures were sampled from MD simulations at 300-10000 K starting from random structures with various densities.
The random structures were obtained from the random structure search at external pressures of 0-2500 GPa using unit cells with 8-16 atoms.
The resulting database contains 75332 structures and 3747903 local environments. Among the dataset structures, randomly chosen 52669 structures were used as the training dataset, the remaining structures were used as the testing dataset.
The contents of the training and testing dataset are summarized in Table \ref{table:traindataset} and Table \ref{table:testdataset}, respectively.

To obtain the reference energies and forces, the spin-polarized DFT calculations were performed using the generalized gradient approximation with the PBE functional \cite{PhysRevLett.77.3865,PhysRevLett.78.1396} and the projector-augmented wave (PAW) method \cite{PhysRevB.50.17953} implemented in VASP \cite{PhysRevB.47.558,Kresse_1994,PhysRevB.54.11169,PhysRevB.59.1758}. The PREC setting in VASP, which determines the energy cutoff and the fast Fourier transform grids, was set to Accurate, leading to a 400~eV cutoff. A $\Gamma$-centered k-point mesh with 0.033~\r{A}$^{-1}$ spacing was used for the periodic directions, and one k-point for the vacuum directions. The Gaussian smearing was used with a 0.05~eV smearing width. The convergence criterion for the electronic self-consistent loop was set to 10$^{-4}$~eV. In the following sections, unless otherwise noted, the settings for DFT calculations are the same as those described here.

\begin{table}[phtb!]
\caption{\label{table:traindataset}
Contents of our training dataset. The first column is the structure type. The second column is the number of structures. The third column is the number of distinct local environments (atoms).}
\begin{ruledtabular}
\begin{tabular}{lcc}
 Structure type & No. structures & No. environments \\
\midrule
$sp^2$ structure && \\ 
\hspace{3mm}Monolayer graphene & 4791 & 499388 \\
\hspace{3mm}AA-bilayer graphene & 560 & 35840 \\
\hspace{3mm}AB-bilayer graphene & 560 & 35840 \\
\hspace{3mm}AA-graphite & 560 & 35840 \\
\hspace{3mm}AB-graphite & 560 & 35840 \\
\hspace{3mm}Nanotube & 2779 & 268096 \\
Diamond & 5052 & 743623 \\
General crystal && \\ 
\hspace{3mm}bc8 & 1061 & 135808 \\
\hspace{3mm}sc & 1112 & 139000 \\
\hspace{3mm}bcc & 840 & 107520 \\
\hspace{3mm}fcc & 840 & 90720 \\
Cluster && \\ 
\hspace{3mm}Ring & 2519 & 18440 \\
\hspace{3mm}Chain & 4200 & 15120 \\
\hspace{3mm}Fullerene & 910 & 33662 \\
Disordered & 9340 & 340136 \\
Random & 16985 & 164720 \\
\midrule
Total & 52669 & 2699593 \\
\end{tabular}
\end{ruledtabular}
\end{table}

\begin{table}[phtb!]
\caption{\label{table:testdataset}
Contents of our testing dataset. The columns are the same as in Table \ref{table:traindataset}.}
\begin{ruledtabular}
\begin{tabular}{lcc}
 Structure type & No. structure & No. environment \\
\midrule
$sp^2$ structure && \\ 
\hspace{3mm}Monolayer graphene & 2184 & 220950 \\
\hspace{3mm}AA-bilayer graphene & 240 & 15360 \\
\hspace{3mm}AB-bilayer graphene & 240 & 15360 \\
\hspace{3mm}AA-graphite & 240 & 15360 \\
\hspace{3mm}AB-graphite & 240 & 15360 \\
\hspace{3mm}Nanotube & 1190 & 115544 \\
Diamond & 1401 & 247121 \\
General crystal && \\ 
\hspace{3mm}bc8 & 237 & 30336 \\
\hspace{3mm}sc & 240 & 30000 \\
\hspace{3mm}bcc & 360 & 46080 \\
\hspace{3mm}fcc & 360 & 38880 \\
Cluster && \\ 
\hspace{3mm}Ring & 1079 & 7934 \\
\hspace{3mm}Chain & 1800 & 6480 \\
\hspace{3mm}Fullerene & 390 & 14338 \\
Disordered & 4003 & 146151 \\
Random & 8459 & 83056 \\
\midrule
Total & 22663 & 1048310 \\
\end{tabular}
\end{ruledtabular}
\end{table}

\subsection{Training results}

We trained MLBOP on our dataset using various hyperparameter settings, including the dimension of $\bm{\zeta}_{ij}$, the architectures of $\phi_{M}$, $\phi_{R}$, and $\phi_{A}$, and the cutoff distance $R_{c}$. For comparison, we also trained other MLIP models, DeePMD~\cite{PhysRevLett.120.143001}, SchNet~\cite{10.1063/1.5019779} and ACE \cite{PhysRevB.99.014104}, on our training dataset. 

For the training of DeePMD, we used DeePMD-kit \cite{WANG2018178}. The local environments were represented using the Deep Potential Smooth Edition, which was introduced in Ref. \cite{zhang2018endtoendsymmetrypreservinginteratomic} and implemented as the se\_e2\_a descriptor in the DeePMD-kit. The cutoff distance was set to 4.0~\r{A} to define the local environment, with a smoothing cutoff of 0.5~\r{A}. The embedding net consists of 3 hidden layers, each having 25, 50, and 100 neurons. The number of axis neurons is set to 16. The fitting net has 3 hidden layers with 256 neurons each. The GELU activation function \cite{hendrycks2023gaussianerrorlinearunits} was applied for each hidden layer in the embedding and fitting net. The energy loss and force loss weights were set to 0.1 and 1.0, respectively, and remained constant throughout the training. The training was performed using the ADAM optimizer for 20000000 batches with a batch size of 4. The initial learning rate was set to 10$^{-3}$ and exponentially decayed to 10$^{-6}$ by the end of the training.

For the training of SchNet, we used the implementation in the documentation of PyTorch Geometric \cite{pyg_documentation} with modifications for the force calculation in periodic boundary conditions. The cutoff distance was set to 4.0~\r{A} for each interaction block. The number of interaction blocks was 3. The number of feature dimensions, filters, and Gaussians were set to 128, 128, and 40, respectively.
The loss function was computed, as shown in Eq. \ref{eq:lossfunc}, with $\rho = 0.1$.
The training was performed with the mini-batch stochastic gradient descent using the ADAM optimizer with a mini-batch size of 4. The learning rate was initially set to $1.0\times10^{-3}$, which exponentially decayed~every epochs with a decaying rate of 0.99. The training was continued until the learning rate reached $1.0\times10^{-6}$. 

For the training of ACE, we used pacemaker \cite{lysogorskiy2021performant}. The cutoff distance was set to 4.0~\r{A}. The exponentially-scaled Chebyshev polynomials were used for the radial basis functions. A total of 603 basis functions up to the fifth body order were applied. The loss weight parameter $\kappa$ was set to 0.6. The loss function was calculated only from energy and force losses without regularization, and the uniform weighting policy was used to assign equal weights to all structures. Training was performed using the BFGS algorithm with the ladder basis extension. In the last ladder step, the training was continued until it was automatically stopped by the early stopping function. For the early stopping, the minimal relative change in the train loss per iteration was set to $1.0\times10^{-5}$ with a patience of 200 steps.

Table \ref{table:hyperparameter} summarizes the mean absolute errors (MAEs) for energy and force on both the training and testing datasets, comparing MLBOP with various hyperparameter settings (different MLP sizes and cutoff distances) against DeePMD, SchNet, and ACE. The CPU times were measured by performing MD simulations of liquid carbon with a density of 2.0 g/cm$^3$ at 4000 K, following the method in Ref. \cite{doi:10.1021/acs.jctc.2c01149} for consistency.
Increasing the dimension of the vector $\zeta_{ij}$ and the sizes of the MLPs $\phi_{M}$, $\phi_{R}$, and $\phi_{A}$ consistently improves the MAEs for both energy and force predictions. This improvement in accuracy, however, comes with a increase in computational cost. This shows a trade-off between model accuracy and computational efficiency. When the network architecture is fixed, varying the cutoff radius among 3.6, 4.0, and 4.4~\r{A} shows that a cutoff of 4.0~\r{A} provides the most accurate fit. 

Overall, MLBOP achieves lower energy and force MAEs than other models while keeping a smaller number of parameters. Even with very compact MLPs, MLBOP can represent a wide range of potential energy surfaces with reasonable accuracy. This indicates that the MLBOP architecture effectively captures the global features of the potential energy surface without relying solely on the expressive power of the MLP, which is essential for ensuring transferability across diverse configurations.

At the same time, despite its compactness, the model still requires a moderate amount of CPU time. One possible reason for this is the $O(N_\mathrm{total} \times N_\mathrm{neigh}^2)$ scaling in the computation of $E_\mathrm{bond}$, where $N_\mathrm{total}$ is the total number of atoms in the system and $N_\mathrm{neigh}$ is the number of neighboring atoms, which is inherited from the original Tersoff potential formulation. This computational cost can be reduced to $O(N_\mathrm{total} \times N_\mathrm{neigh})$ by reformulating the angular dependence, for example, using spherical harmonics and Clebsch–Gordan contractions, as demonstrated in Refs. \cite{PhysRevLett.55.2001,PhysRevB.87.184115,PhysRevB.99.014104,Batzner2022}. In addition, more efficient matrix operation implementations may further improve performance. Although computational speed was not the primary focus of this study, these directions remain important considerations for future work.

\begin{figure*}[bt]
\centering
\includegraphics[clip,scale=0.37]{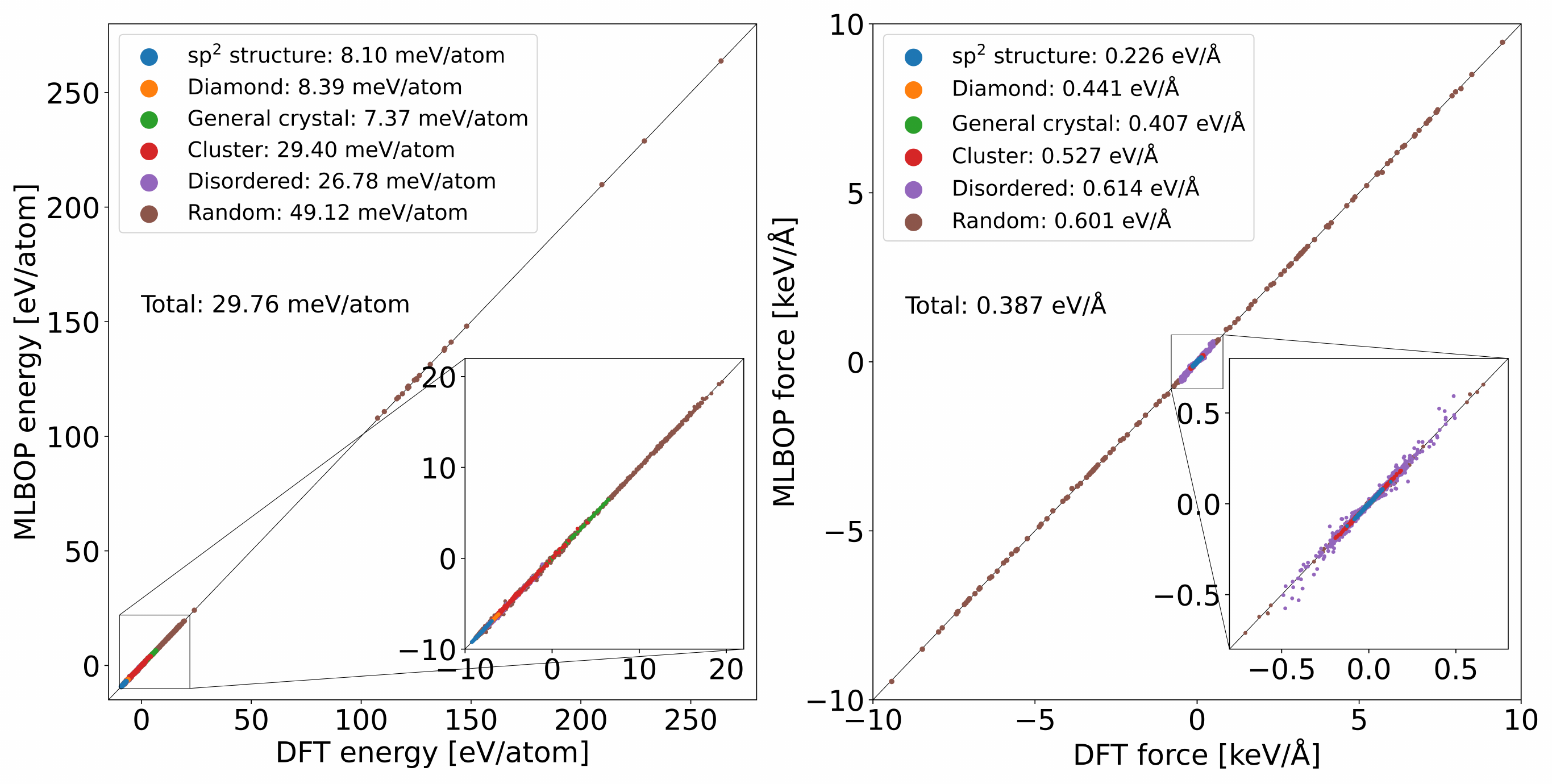}
\caption{Comparison between DFT and MLBOP for the testing dataset. Top: Energy. Bottom: Force. The colors of the circles represent the type of the dataset structure. The MLBOP energies are adjusted by a constant shift to match the DFT energies. Insets show magnified views for the low energy regions. The values in the panels are the mean absolute errors (MAEs) for each structure type.}
\label{fig:eneparity}
\end{figure*}

The following section \ref{sec:verification} presents validation results for MLBOP trained with a cutoff length $R_{c}$ of 4.0~\r{A}, a vector size $\bm{\zeta}_{ij}$ of 20, network structures of 20-20-20-1 for $\phi_{R}$ and $\phi_{A}$, and 3-5-10-20 for $\phi_{M}$. Using these settings, which are listed in the third row of Table \ref{table:hyperparameter}, MLBOP achieves the MAEs of 29.76~meV/atom for energy and 0.387~eV/\r{A} for force on the testing dataset, and shows good agreement with DFT across a wide range of the potential energy surface, as shown in Figure \ref{fig:eneparity}.

\begin{table*}[bt]
\caption{\label{table:hyperparameter}
MAEs of energies $E$ and forces $F$ on our training and testing datasets, and CPU time of MLBOP with different hyperparameters, compared to DeePMD, SchNet, and ACE. The architecture of MLPs $\phi_{M}$, $\phi_{R}$, and $\phi_{A}$ is described using a hyphenated sequence, where each number corresponds to the number of units from the input layer to the output layer. The MAEs are given in~meV/atom for energy and~eV/\r{A} for force. The CPU time is given in $\mu$s/atom/step. The timings were performed using the Atomic Simulation Environment (ASE) \cite{ase-paper} for SchNet, and LAMMPS for MLBOP, ACE, and DeePMD, on a single Intel Xeon Platinum 8360Y 2.4 GHz CPU core.}
\begin{ruledtabular}
\begin{tabular}{llccccccccc}
& $\bm{\zeta}_{ij}$ & $\phi_{M}$ & $\phi_{R}$ and $\phi_{A}$ & No. parameters & $R_{c}$ & $E_\mathrm{train}$ & $F_\mathrm{train}$ & $E_\mathrm{test}$ & $F_\mathrm{test}$ & CPU time \\ 
\midrule
& 8 &3-3-4-8 & 8-8-8-1 & 395 & 4.0 & 71.95 & 0.592 & 74.72 & 0.586 & 424 \\
 & 12 &3-3-6-12 & 12-12-12-1 & 791 & 4.0 & 56.70 & 0.512 & 58.78 & 0.504 & 672\\
 & 20\footnotemark[1] & 3-5-10-20 & 20-20-20-1 & 2047 & 4.0 &27.81 & 0.395 & 29.76 & 0.387 & 1631 \\
MLBOP & 28 & 3-7-14-28 & 28-28-28-1 & 3895 & 4.0 & 25.83 & 0.381 & 27.54 & 0.376 & 3333\\
 & 36 & 3-9-18-36 & 36-36-36-1 & 6335 & 4.0 & 25.04 & 0.378 & 26.94 & 0.375 & 4098\\
& 20 & 3-5-10-20 & 20-20-20-1 & 2047 & 3.6 & 37.83 & 0.451 & 40.04 & 0.439 & 917 \\
& 20 & 3-5-10-20 & 20-20-20-1 & 2047 & 4.4 & 36.62 & 0.454 & 37.82 & 0.421 & 2724 \\
\midrule
DeePMD & \-- & \-- & \-- & 550260 & 4.0 & 89.41 & 0.584  & 92.78 & 0.581 & 1104 \\
SchNet & \-- & \-- & \-- & 221953 & 4.0 & 25.54 & 0.492 & 31.42 & 0.510 & 473 \\
ACE & \-- & \-- & \-- & 3002 & 4.0 & 145.9 & 0.374 & 140.3 & 0.399 & 91 \\
\end{tabular}
\end{ruledtabular}
\footnotetext[1]{MLBOP trained with the hyperparameters in this row is validated in section V.}
\end{table*}

\clearpage

\section{Validation}
\label{sec:verification}
In developing interatomic potentials, regression metrics on a given test dataset are insufficient to measure performance. To confirm the applicability of an interatomic potential, it is essential to evaluate it from multiple perspectives. In this section, we validate the MLBOP developed in section \ref{sec:training} by checking its accuracy on various physical properties.

\subsection{Crystal}
\begin{figure}[phtb!]
\centering
\includegraphics[clip,scale=0.39]{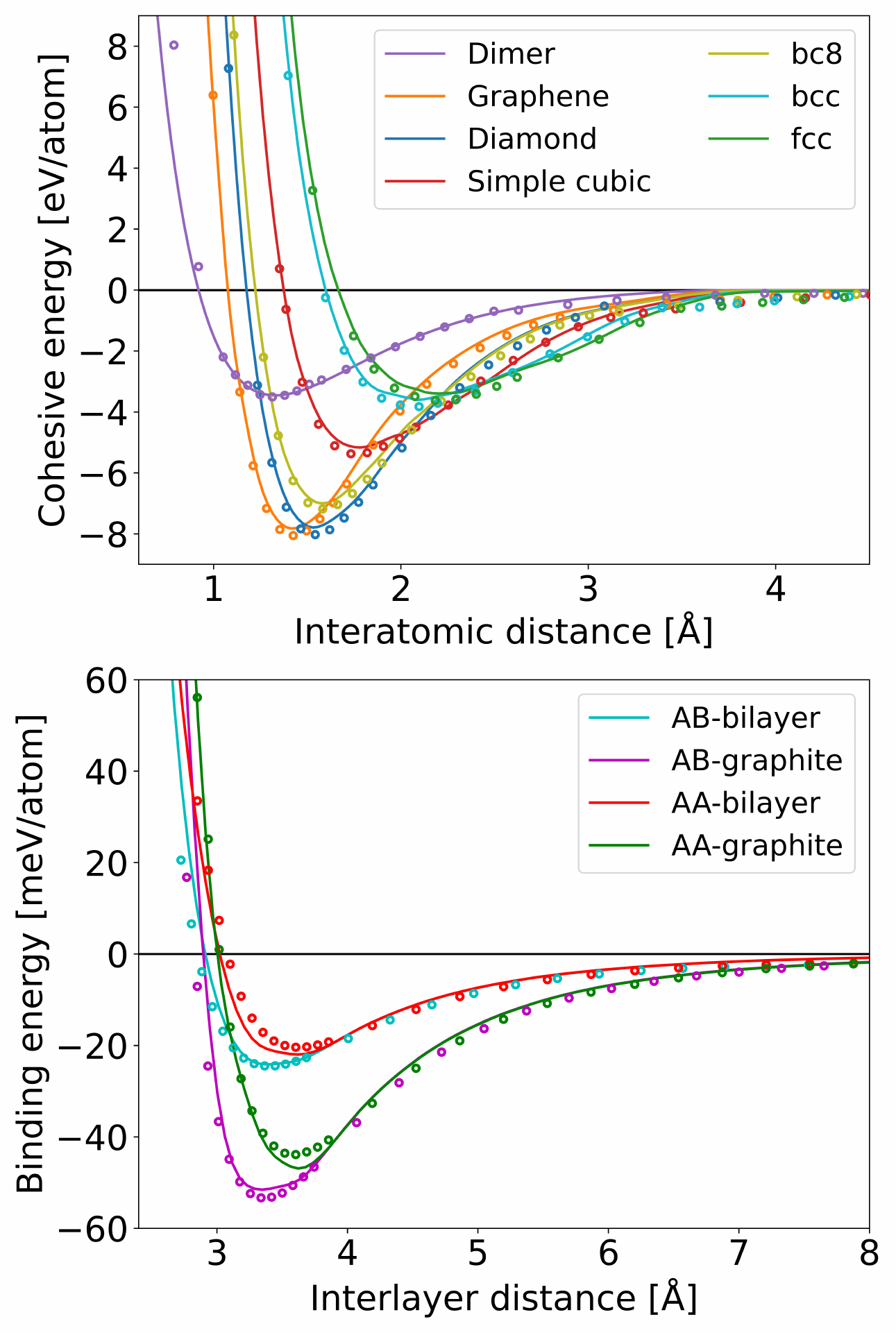}
\caption{Top: Cohesive energy vs nearest-neighbor interatomic distance curves for dimer, monolayer graphene, diamond, simple cubic, bc8, body-centered cubic (bcc), and face-centered cubic (fcc) calculated with MLBOP and DFT (PBE-D3). Bottom: Binding energy vs interlayer distance curves for AB- and AA-stacked graphite and bilayer graphene calculated with MLBOP and DFT (PBE-D3). The lines indicate the MLBOP energies and the circles indicate the DFT energies.}
\label{fig:potentialene}
\end{figure}

As a first test, we calculated the cohesive energy vs the nearest neighbor interatomic distance curves using DFT and MLBOP for dimer, monolayer graphene, diamond, simple cubic, bc8, bcc, and fcc, as shown in the top in Figure \ref{fig:potentialene}. The cohesive energy curves calculated with MLBOP are in good agreement with the DFT and have good smoothness over a wide range of interatomic distances. We also calculated the binding energy of AA- and AB-stacked graphite and bilayer graphene as a function of the interlayer distance using DFT and MLBOP, as shown in the bottom in Figure \ref{fig:potentialene}, and the binding energy curves calculated with MLBOP also agree with DFT. 

\begin{figure*}[phtb!]
\centering
\includegraphics[clip,scale=0.18]{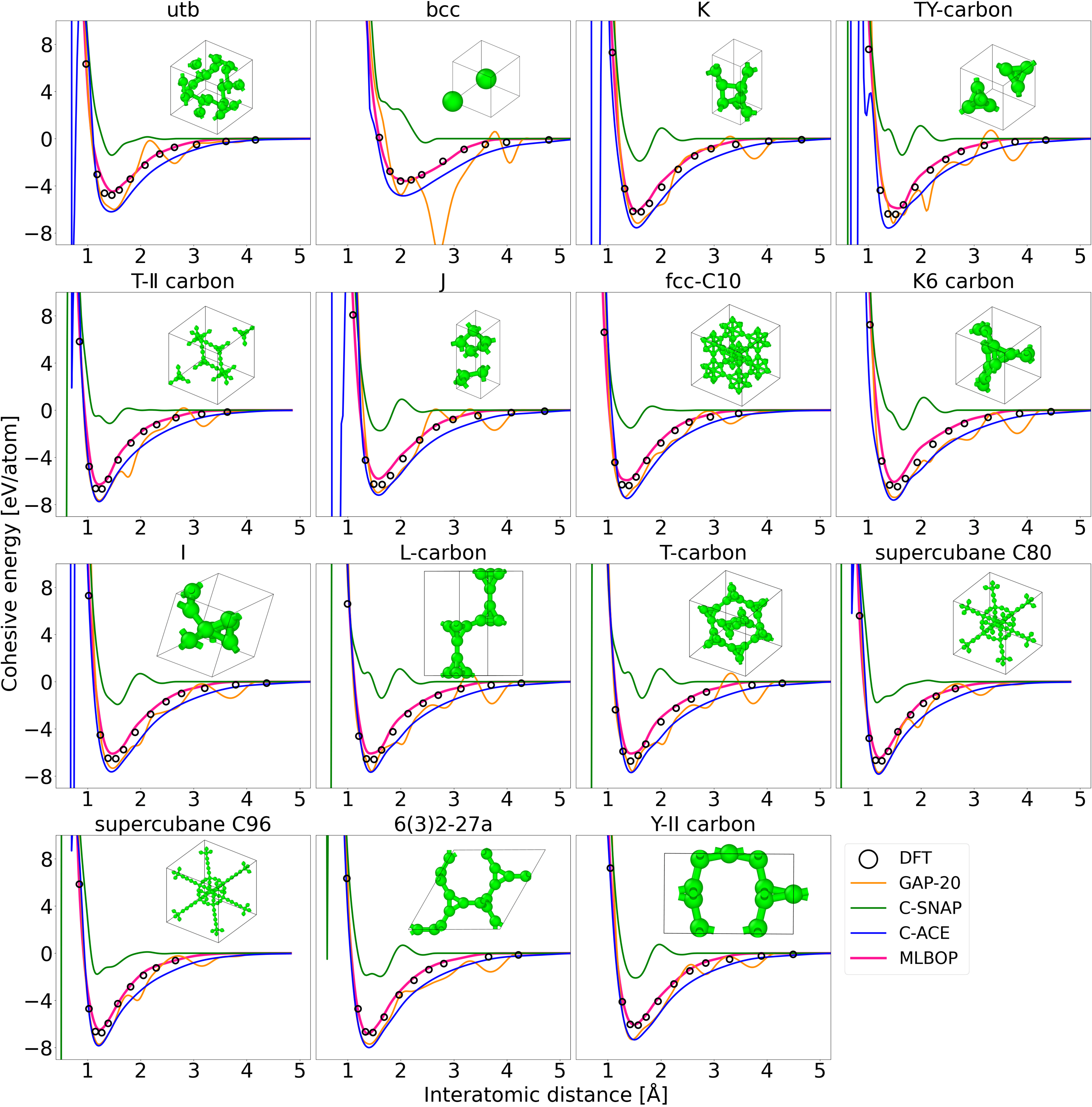}
\caption{Cohesive energy vs nearest-neighbor interatomic distance curves for the unstable crystals in the SACADA database, calculated with DFT (PBE-D3), GAP-20, C-SNAP, C-ACE, and MLBOP. The unit cells of the crystals are shown in each panel.}
\label{fig:scadaene}
\end{figure*}

In addition, for unbiased validation, the cohesive energy curves of the first to fifteenth most unstable crystals in the SACADA database, which are not explicitly included in the training dataset except for bcc, were computed again.
Figure \ref{fig:scadaene} shows the cohesive energy curves calculated using DFT (PBE-D3), MLBOP, GAP-20 \cite{10.1063/5.0005084,10.1063/5.0091698}, C-SNAP \cite{PhysRevB.106.L180101}, and C-ACE \cite{doi:10.1021/acs.jctc.2c01149}.
The MLBOP curves have no unphysical local minima and good smoothness in agreement with the DFT over a wider range of energy curves compared to other reported MLIPs.

\subsection{Defect}
\label{sec:defect}
We evaluate MLBOP by calculating defect formation energies for various defects in monolayer graphene, diamond, (5,5) and (9,0) nanotubes. The defect formation energy $E_\mathrm{f}$ is defined as
\begin{equation} 
E_\mathrm{f} = E_\mathrm{d} - N\mu,
\label{eq:defect}
\end{equation}
where $E_\mathrm{d}$ is the total energy of the defective structure, $N$ is the number of atoms in the defective structure, and $\mu$ is the per-atom energy of the defect-free structure. Each defective structure was generated with a supercell of (11$\times$11$\times$1) with 242 atoms for graphene, (3$\times$3$\times$3) with 216 atoms for diamond, supercells with 180 atoms and 178 atoms for (5,5) and (9,0) nanotubes, respectively. The defect-free structures were relaxed with respect to the atomic configurations and cell parameters. The defective structures were generated by manipulating atoms in the supercell of the defect-free structure, followed by relaxation to local minima while keeping cell parameters fixed. The structural optimizations and defect formation energy calculations were performed using DFT, MLBOP, the second generation REBO potential (REBO-\RomanNumeralCaps{2}) \cite{DonaldWBrenner_2002}, GAP-20, C-SNAP, and C-ACE. The DFT calculations were performed using two exchange correlation functionals, PBE-D3 and optB88 \cite{PhysRevB.82.081101,PhysRevB.83.195131,PhysRevLett.92.246401,Klimeš_2010}. Although some defect formation energies have been previously reported, all values were recalculated to ensure consistency in computational conditions and protocols.
For the DFT calculations, the convergence criteria for the electronic self-consistent loop and the ionic relaxation loop were set to 10$^{-6}$ and 10$^{-5}$~eV, respectively. 
For the interatomic potentials, the structural optimizations were performed using LAMMPS \cite{LAMMPS}, and the convergence criteria of energy and force were set to 10$^{-10}$~eV and 10$^{-8}$~eV/\r{A}, respectively.
Table \ref{table:defect} lists the calculated defect formation energies. 
MLBOP predicted the defect formation energies with accuracies comparable to or better than other MLIPs.

\begin{table*}[htb!]
\caption{\label{table:defect}
Formation energies of various defects in graphene, nanotubes, and diamond. The formation energies were calculated with DFT (PBE-D3 and optB88), MLBOP, REBO-\RomanNumeralCaps{2}, GAP-20, C-SNAP, and C-ACE. In brackets is the absolute percentage error from the PBE-D3 value. The bottom row shows the mean absolute percentage error for each method.}
\begin{ruledtabular}
\begin{tabular}{lccccccc}
& \multicolumn{7}{c}{Defect formation energy [eV] (\% error)} \\
\cmidrule{2-8}
 & DFT & \multirow{2}{*}{MLBOP} & \multirow{2}{*}{REBO-\RomanNumeralCaps{2}} & \multirow{2}{*}{GAP-20} & \multirow{2}{*}{C-SNAP} & \multirow{2}{*}{C-ACE} & DFT \\
 & (PBE-D3) &  &  &  &  &  & (optB88)\\
\midrule
Graphene Stone-Wales & 4.79 & 4.70 (2) & 5.46 (14) & 5.12 (7) & 4.65 (3) & 4.86 (1) & 5.01\\
Graphene monovacancy & 7.82 & 7.24 (7) & 7.52 (4) & 6.99 (11) & 3.19 (59) & 7.05 (10) & 8.03\\
Graphene divacancy (5-8-5) & 7.11 & 7.27 (2) & 7.37 (4) & 7.86 (11) & 6.08 (14) & 7.22 (2) & 7.29\\
Graphene divacancy (555-777) & 6.54 & 6.08 (7) & 6.68 (2) & 6.87 (5) & 6.72 (3) & 6.35 (3) & 6.78\\
Graphene divacancy (5555-6-7777) & 6.64 & 6.36 (4) & 7.38 (11) & 7.37 (11) & 7.68 (16) & 6.89 (4) & 6.87\\
Graphene adatom & 6.48 & 5.55 (14) & 6.09 (6) & 5.26 (19) & -2.04 (131) & 5.29 (18) & 6.40\\ \midrule
(5,5) nanotube monovacancy & 6.75 & 6.28 (7) & 6.28 (7) & 6.36 (6) & -0.41 (106) & 5.67 (16) & 6.73\\
(5,5) nanotube divacancy & 4.38 & 3.91 (11) & 5.01 (14) & 4.58 (5) & -4.50 (203) & 4.35 (1) & 4.33\\
(5,5) nanotube Stone-Wales (parallel) & 3.78 & 3.38 (11) & 4.80 (27) & 3.88 (3) & -5.17 (237) & 3.74 (1) & 3.71\\
(5,5) nanotube Stone-Wales (transverse) & 3.12 & 2.66 (15) & 3.67 (18) & 3.06 (2) & -2.25 (172) & 3.07 (2) & 3.11\\ \midrule
(9,0) nanotube monovacancy & 5.54 & 4.92 (11) & 4.90 (12) & 4.92 (11) & -2.99 (154) & 4.52 (18) & 5.52\\
(9,0) nanotube divacancy & 3.38 & 3.27 (3) & 4.25 (26) & 3.53 (4) & -3.54 (205) & 3.42 (1) & 3.39\\
(9,0) nanotube Stone-Wales (parallel) & 2.91 & 2.71 (7) & 3.56 (23) & 2.93 (1) & -0.06 (102) & 3.09 (6) & 2.92\\
(9,0) nanotube Stone-Wales (transverse) & 3.68 & 3.11 (16) & 4.37 (19) & 3.33 (10) & -3.41 (193) & 3.54 (4) & 3.69 \\ \midrule
Diamond monovacancy & 7.01 & 6.70 (4) & 7.17 (2) & 4.28 (39) & 1.48 (79) & 5.73 (18) & 6.55\\
Diamond divacancy & 9.51 & 10.01 (5) & 10.77 (13) & 6.71 (29) & 2.23 (77) & 8.93 (6) & 8.98\\
Diamond interstitial & 11.86 & 10.26 (13) & 10.05 (15) & 7.98 (33) & 8.76 (26) & 8.42 (29) & 11.47 \\ 
\midrule
Mean absolute percentage error [\%] & \-- & 8.2 & 12.7 & 12.1 & 104.6 & 8.2 & \--
\end{tabular}
\end{ruledtabular}
\end{table*}

\subsection{Surface}
We evaluate MLBOP by calculating surface energies for diamonds with various surface orientations.
The surface energy $E_\mathrm{surf}$ is calculated as
\begin{equation} 
E_\mathrm{surf} = \frac{E_\mathrm{slab} - N\mu}{2A},
\label{eq:surface}
\end{equation}
where $E_\mathrm{slab}$ is the total energy of the surface slab, $N$ is the number of atoms in the slab, $\mu$ is the per-atom energy of the bulk structure, and $A$ is the slab surface area.
We calculated the surface energies of both ideal and relaxed surfaces with (100), (111), and (110) orientations, as well as the (2$\times$1)-reconstructed surfaces of (100) and (111), called Pandey-chain reconstruction.
The reconstructed surfaces were obtained from DFT-MD simulations of diamond slabs, followed by the optimization of atomic configurations to local energy minima using the other interatomic potentials tested. The settings for the structural optimizations were the same as those used for the defect formation energy calculations.
Table \ref{table:surface} shows the surface energies calculated with DFT (PBE-D3, optB88, and B3LYP \cite{10.1063/1.464913,PhysRevB.37.785,doi:10.1021/j100096a001}), MLBOP, REBO-\RomanNumeralCaps{2}, GAP-20, C-SNAP, and C-ACE. MLBOP achieves a similar level of accuracy in predicting the surface energies as GAP-20 and C-ACE.
We also calculated the edge energies of graphene and carbon nanotubes with different edge orientations: zigzag and armchair. The edge energy $E_\mathrm{edge}$ is defined as
\begin{equation} 
E_\mathrm{edge} = \frac{E_\mathrm{cut} - N\mu}{2L},
\label{eq:edge}
\end{equation}
where $E_\mathrm{cut}$ is the total energy of the structure with the exposed edge, $N$ is the number of atoms in this structure, $\mu$ is the per-atom energy of the edge-free structure, and $L$ is the length of the edge. Table \ref{table:edge} shows the zigzag edge energy of graphene and a (9,0) nanotube and the armchair edge energy of graphene and a (5,5) nanotube. For both graphene and nanotubes, DFT calculation predicts that the armchair edge energy is lower than the zigzag edge energy. MLBOP can reproduce this trend, as can GAP-20 and C-ACE. 

\begin{table*}[htb!]
\caption{\label{table:surface}
Surface energies of diamond with different orientations calculated with DFT (PBE-D3, optB88, and B3LYP), MLBOP, REBO-\RomanNumeralCaps{2}, GAP-20, C-SNAP, and C-ACE. The B3LYP values are from Ref. \cite{doi:10.1080/00268976.2013.829250}. In brackets is the absolute percentage error from the PBE-D3 value. The bottom row shows the mean absolute percentage error for each method.}
\begin{ruledtabular}
\begin{tabular}{lcccccccc}
& \multicolumn{8}{c}{Surface energy [eV/\r{A}$^2$] (\% error)} \\
\cmidrule{2-9}
 & DFT & \multirow{2}{*}{MLBOP} & \multirow{2}{*}{REBO-\RomanNumeralCaps{2}} & \multirow{2}{*}{GAP-20}  & \multirow{2}{*}{C-SNAP} & \multirow{2}{*}{C-ACE} & DFT & DFT \\
 & (PBE-D3) &  &  &  &  &  & (optB88) & (B3LYP) \\
\midrule
(100) 1$\times$1 ideal & 0.579 & 0.634 (10) & 0.727 (26) & 0.641 (11) & 0.221 (62) & 0.590 (2) & 0.557 & 0.582\\
(100) 1$\times$1 relaxed & 0.562 & 0.603 (7) & 0.722 (29) & 0.586 (4) & 0.212 (62) & 0.573 (2) & 0.539 & 0.567\\
(100) 2$\times$1 reconstructed & 0.341 & 0.352 (3) & 0.435 (28) & 0.322 (6) & 0.163 (52) & 0.382 (12) & 0.322 & 0.301\\
(111) 1$\times$1 ideal & 0.481 & 0.457 (5) & 0.465 (3) & 0.371 (23) & 0.092 (81) & 0.514 (7) & 0.483 & 0.506\\
(111) 1$\times$1 relaxed & 0.423 & 0.411 (3) & 0.423 (0) & 0.263 (38) & 0.084 (80) & 0.405 (4) & 0.422 & 0.403\\
(111) 2$\times$1 reconstructed & 0.239 & 0.272 (14) & 0.234 (2) & 0.224 (7) & 0.224 (6) & 0.315 (32) & 0.239 & 0.235\\
(110) 1$\times$1 ideal & 0.453 & 0.405 (11) & 0.289 (36) & 0.340 (25) & 0.097 (79) & 0.451 (0) & 0.439 & 0.465\\
(110) 1$\times$1 relax & 0.367 & 0.357 (3) & 0.220 (40) & 0.258 (30) & 0.081 (78) & 0.374 (2) & 0.346 & 0.406\\
\midrule
Mean absolute percentage error [\%] & \-- & 6.9 & 20.5 & 17.9 & 62.5 & 7.6 & \-- & \-- 
\end{tabular}
\end{ruledtabular}
\end{table*}

\begin{table*}[htb!]
\caption{\label{table:edge}
Edge energies of graphene and nanotubes calculated with DFT (PBE-D3 and optB88), MLBOP, REBO-\RomanNumeralCaps{2}, GAP-20, C-SNAP, and C-ACE. In brackets is the absolute percentage error from the PBE-D3 value. The bottom row shows the mean absolute percentage error for each method.}
\begin{ruledtabular}
\begin{tabular}{lccccccc}
& \multicolumn{7}{c}{Edge energy [eV/\r{A}] (\% error)} \\
\cmidrule{2-8}
 & DFT & \multirow{2}{*}{MLBOP} & \multirow{2}{*}{REBO-\RomanNumeralCaps{2}} & \multirow{2}{*}{GAP-20} & \multirow{2}{*}{C-SNAP} & \multirow{2}{*}{C-ACE} & DFT \\
 & (PBE-D3) &  &  &  &  &  & (optB88) \\
\midrule
Graphene armchair & 1.04 & 1.03 (1) & 1.09 (5) & 1.11 (7) & 0.47 (55) & 0.94 (10) & 1.02\\
Graphene zigzag & 1.20 & 1.16 (4) & 1.04 (14) & 1.25 (3) & 0.39 (68) & 0.97 (20) & 1.23\\
(5,5) nanotube armchair & 0.96 & 1.01 (6) & 1.01 (6) & 1.00 (5) & 0.37 (61) & 0.88 (8) & 0.93\\
(9,0) nanotube zigzag & 1.18 & 1.15 (3) & 0.96 (19) & 1.04 (12) & 0.29 (75) & 0.93 (21) & 1.20\\
\midrule
Mean absolute percentage error [\%] & - & 3.2 & 10.7 & 6.8 & 64.8 & 14.5 & - 
\end{tabular}
\end{ruledtabular}
\end{table*}

\subsection{Liquid}
\begin{figure}[htbp!]
\centering
\includegraphics[clip,scale=0.26]{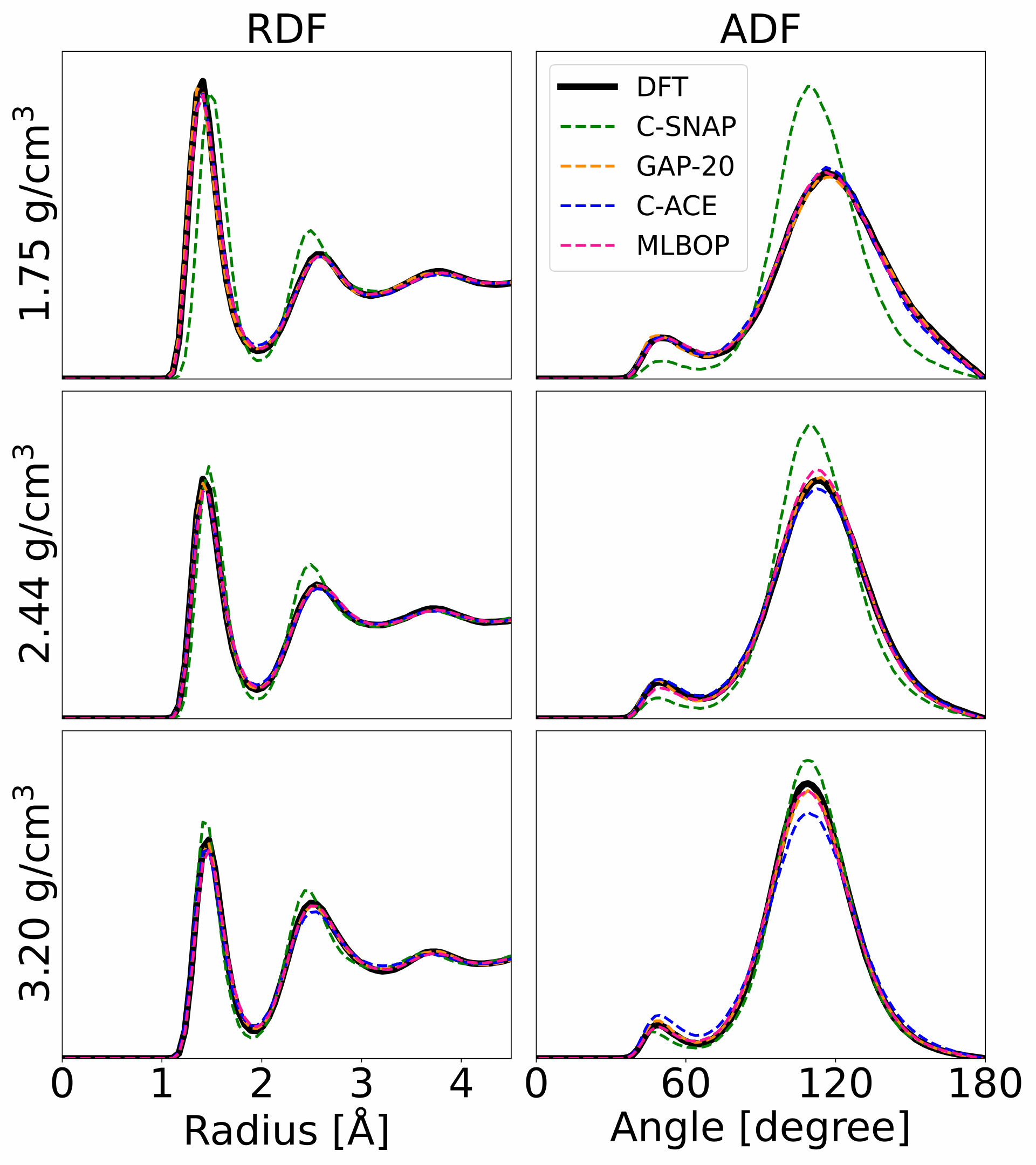}
\caption{Radial distribution functions (RDFs) and angular distribution functions (ADFs) of liquid carbon with densities of 1.75 (top row), 2.44 (middle row), and 3.20 g/cm$^3$ (bottom row) at 5000 K for DFT (PBE-D3), C-SNAP, GAP-20, C-ACE, and MLBOP.}
\label{fig:liquid}
\end{figure}

We compared liquid structures from molecular dynamics (MD) simulations using DFT, MLBOP, C-SNAP, GAP-20, and C-ACE. The DFT-based MD simulations employed the PBE-D3 functional with a single k-point and were performed using VASP with the PREC setting set to Medium. Each simulation included 64 atoms at densities of 1.75, 2.44, and 3.20 g/cm$^3$, and was run for 30 ps at 5000 K with a 2.0 fs timestep, starting from a simple cubic configuration. Radial distribution functions (RDF) and angular distribution functions (ADF) were averaged over the final 15 ps of each trajectory. Figure~\ref{fig:liquid} presents the RDF and ADF at each density. MLBOP closely reproduces the DFT results across all distances and angles without introducing spurious peaks, similar to GAP-20 and C-ACE. While C-SNAP shows good agreement with DFT at the highest density of 3.20 g/cm$^3$, it exhibits discrepancies from DFT at lower densities.

\subsection{Amorphous phase}
\begin{figure}[htbp]
\centering
\includegraphics[clip,scale=0.40]{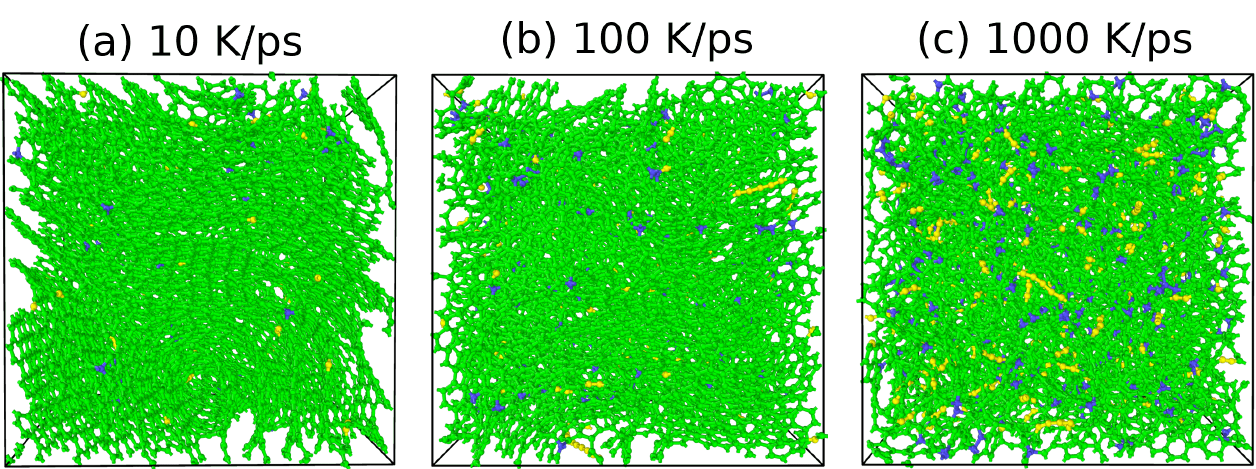}
\includegraphics[clip,scale=0.53]{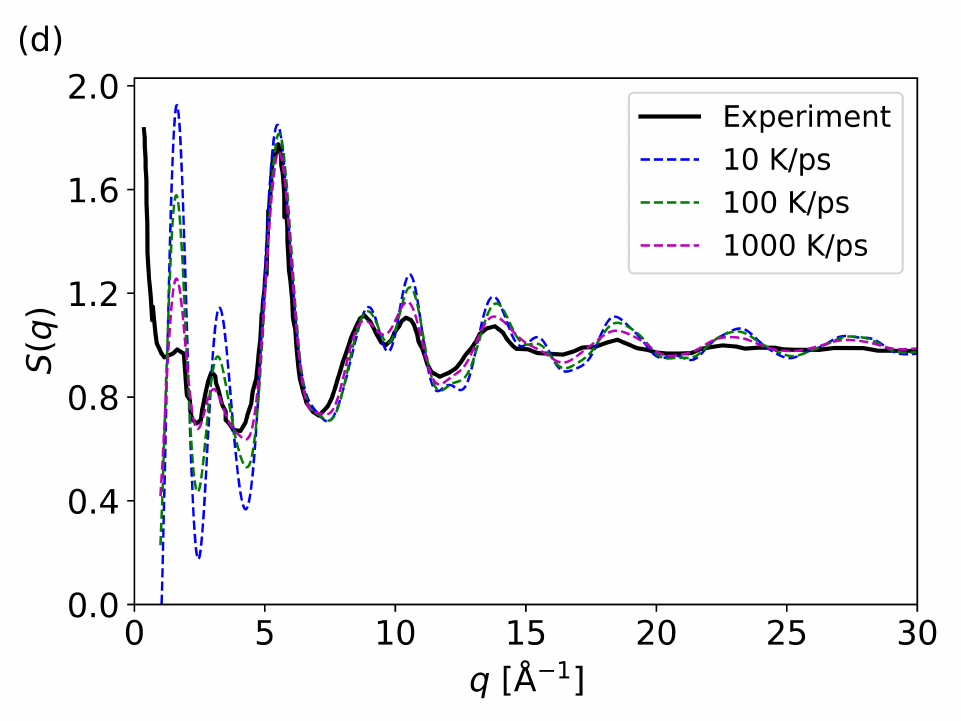}
\caption{Amorphous carbon with a density of 2.0 g/cm$^3$ obtained from the melt-quench MD simulations of 8000 atoms at cooling rates of (a) 10 K/ps, (b) 100 K/ps, and (c) 1000 K/ps. The yellow, green, and blue balls represent two-, three-, and four-coordinated carbon atoms, respectively. The cutoff distance for calculating the coordination number is 1.85~\r{A}. (d) 
Comparison of the structure factor $S(q)$ between the melt-quench MD simulations and the neutron diffraction of amorphous carbon with a macroscopic density of 2.0 g/cm$^3$ \cite{PhysRevLett.65.1905}.}
\label{fig:amorphous}
\end{figure}

We verified MLBOP for the structure of amorphous carbon obtained by melt-quench MD simulation by comparison with the experimental data. For the melt-quench simulations, we generated simple cubic structures with 8000 atoms at a density of 2.0 g/cm$^3$. These structures were randomized by MD simulations at 12000 K for 4 ps, held at 8000 K for 10 ps, and then quenched to 300 K by linearly decreasing temperature at cooling rates of 10 K/ps, 100 K/ps, and 1000 K/ps. The timestep was set to 1.0 fs. Figure \ref{fig:amorphous} compares the structure factor $S(q)$ of amorphous carbon obtained from the melt-quench simulations with that from the neutron diffraction of amorphous carbon film prepared by sputtering \cite{PhysRevLett.65.1905}. The structure of amorphous carbon depends on the cooling rate, with slower cooling rates yielding more stable graphitic structures. 
A fast cooling rate of 1000 K/ps results in the structure factor in good agreement with the neutron diffraction data, showing that MLBOP can produce a realistic amorphous structure by properly choosing a cooling rate, despite the difference in synthesis method between sputtering and melt quenching. 
%which is consistent with the melt-quench simulations \cite{Jana_2019,doi:10.1021/acs.jctc.2c01149} performed using GAP-17 \cite{PhysRevB.95.094203} and C-ACE, where a fast cooling rate of 1000 K/ps yielded $sp^3$ density profiles that are in better agreement with the experimental results than slower cooling rates.

\subsection{Phonon}
\begin{figure}[phtb!]
\centering
\includegraphics[clip,scale=0.265]{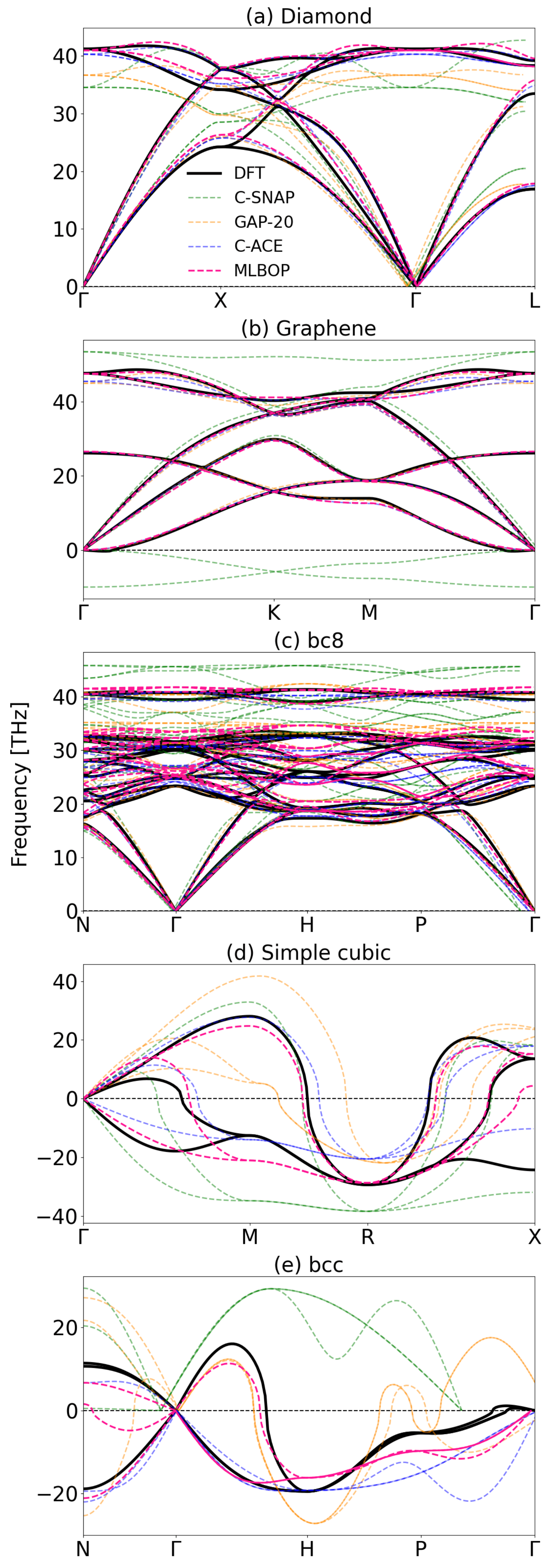}
\caption{Phonon band structures of (a) diamond, (b) graphene, (c) bc8, (d) simple cubic, and (e) bcc calculated with DFT (PBE-D3), GAP-20, C-SNAP, C-ACE, and MLBOP.}
\label{fig:phonon}
\end{figure}

We tested MLBOP on the phonon band structures of diamond, graphene, bc8, and simple cubic.
Phonon calculations were performed using DFT (PBE-D3), GAP-20, C-ACE, C-SNAP, and MLBOP. These calculations were carried out with phonopy \cite{phonopy-phono3py-JPCM,phonopy-phono3py-JPSJ} following full structural relaxation with each method. Figure \ref{fig:phonon} shows the phonon band structures for each structure. The phonon band structures obtained using MLBOP agree with those from DFT, as do the results from GAP-20 and C-ACE.
While C-SNAP can accurately predict the phonon band structure of dense materials such as diamond and bc8, it exhibits large errors when applied to graphene, which is not the intended target of C-SNAP.

\subsection{Elastic properties}
We calculated the elastic properties of several crystals of carbon using DFT (PBE-D3), MLBOP, REBO-\RomanNumeralCaps{2}, GAP-20, C-SNAP, and C-ACE. The elastic properties were calculated by applying normal and shear deformations to each independent direction, relaxing the atomic positions while keeping the cell size and shape, and fitting the stress-strain data, following the method described in Ref. \cite{deJong2015}. Table \ref{table:elastic} shows the independent components of the elasticity tensor of the dynamically and mechanically stable carbon allotropes: diamond, AB-stacked graphite, and bc8. Of the tested models, C-ACE showed the highest accuracy, yielding the lowest average error in predicting elastic properties. Other models overestimated or underestimated some components of the elasticity tensor. While MLBOP successfully reproduces the overall mechanical stability of each structure, some deviations remain in certain components of the elastic tensor, for example, the $C_{13}$ component in AB-stacked graphite. These deviations may be related to the limited number of tilted cell geometries in the training dataset or the absence of stress terms in the loss function. Incorporating such features is expected to further enhance the accuracy of the model.

\begin{table*}[htbp]
\caption{\label{table:elastic}
Elastic properties of AB-stacked graphite, diamond, and bc8. The elastic properties were calculated with DFT (PBE-D3), MLBOP, REBO-\RomanNumeralCaps{2}, GAP-20, C-SNAP, and C-ACE. In brackets is the absolute percentage error from the PBE-D3 value. The bottom row shows the mean absolute percentage error for each method.}
\begin{ruledtabular}
\begin{tabular}{lccccccc}
& \multicolumn{7}{c}{Elastic Properties [GPa]} \\
\cmidrule{2-8}
 & DFT & \multirow{2}{*}{MLBOP} & \multirow{2}{*}{REBO-\RomanNumeralCaps{2}} & \multirow{2}{*}{GAP-20} & \multirow{2}{*}{C-SNAP} & \multirow{2}{*}{C-ACE} & \multirow{2}{*}{Experiment} \\
 & (PBE-D3) &  &  &  &  &  & \\
\midrule
Diamond $C_{11}$ & 1087.8 & 1166.3 (7) & 1062.4 (2) & 975.4 (10) & 1050.9 (3) & 1017.6 (6) &1078.6 \cite{10.1063/1.2975190}\\
Diamond $C_{12}$ & 142.3 & 274.2 (93) & 132.4 (7) & 45.5 (68) & 125.7 (12) & 149.6 (5) &126.63 \cite{10.1063/1.2975190}\\
Diamond $C_{44}$ & 572.3 & 577.8 (1) & 717.7 (25) & 503.1 (12) & 555.8 (3) & 554.5 (3) &577.56 \cite{10.1063/1.2975190}\\
\midrule
AB-stacked graphite $C_{11}$ & 1158.0 & 1108.0 (4) & 921.2 (20) & 1052.5 (9) & 1300.8 (12) & 1027.2 (11) &1109 \cite{PhysRevB.75.153408}\\
AB-stacked graphite $C_{12}$ & 231.5 & 319.1 (38) & 328.2 (42) & 258.5 (12) & 546.9 (136) & 200.7 (13) &139 \cite{PhysRevB.75.153408}\\
AB-stacked graphite $C_{13}$ & -3.5 & 27.3 (881) & 0.0 (100) & 0.1 (103) & 0.0 (100) & -5.0 (42) &0 \cite{PhysRevB.75.153408}\\
AB-stacked graphite $C_{33}$ & 52.7 & 17.7 (66) & 0.0 (100) & 62.1 (18) & -0.0 (100) & 51.7 (2) &38.7 \cite{PhysRevB.75.153408}\\
AB-stacked graphite $C_{44}$ & 5.7 & 11.2 (97) & 0.0 (99) & 27.2 (378) & 0.1 (99) & 6.4 (13) &5.0 \cite{PhysRevB.75.153408}\\
AB-stacked graphite $C_{66}$ & 463.4 & 394.7 (15) & 297.8 (36) & 384.8 (17) & 378.0 (18) & 411.4 (11) &485 \cite{PhysRevB.75.153408}\\
\midrule
bc8 $C_{11}$ & 1265.2 & 1016.0 (20) & 206.7 (84) & 1315.5 (4) & 1315.6 (4) & 1190.3 (6) &-\\
bc8 $C_{12}$ & 59.9 & 68.0 (14) & 482.3 (706) & 351.1 (486) & 362.8 (506) & 147.3 (146) &-\\
bc8 $C_{44}$ & 565.3 & 412.4 (27) & 125.1 (78) & 476.8 (16) & 416.5 (26) & 465.4 (18) &-\\
\midrule
Mean absolute percentage error [\%] & - & 105.2 & 108.3 & 94.3 & 85.0 & 23.1 & -

\end{tabular}
\end{ruledtabular}
\end{table*}

\subsection{Cluster}
To validate MLBOP's ability to predict stable structures of carbon clusters, we optimized the geometries of clusters using a structural optimization method based on the genetic algorithm (GA) \cite{10.7551/mitpress/1090.001.0001}. This approach involved mating, selection, and mutation operations on candidate structures within a population as described below.

Each candidate structure is represented by $N$ lists of atomic coordinates $\bm{x}_{i}$
\begin{equation} 
G = \{\bm{x}_{1},\bm{x}_{2},....,\bm{x}_{N}\}.
\label{eq:cluster}
\end{equation}
The initial population consists of eight candidate structures, each generated by randomly placing $N$ atoms in a sphere of 6.0~\r{A} diameter and then relaxing the atomic configurations to local minima. 
A child structure is produced using the cut-assemble mating operation \cite{PhysRevLett.75.288}.
In this process, two parent structures, $G$ and $G^{'}$, are randomly selected from the population, randomly rotated, aligned so their centers of mass are at the origin, and cut along the XOY plane. A child structure, $G^{''}$, is then generated by assembling the atoms of $G$ which lie above the plane, and the atoms of $G^{'}$ which lie below the plane. If the resulting child structure does not have the correct number of atoms $N$, the parent clusters are translated equal distances in opposite directions, and the process is repeated. After relaxation to the local minimum, the child structure replaces another candidate in the population if its energy is lower.
As the optimization progresses, the population can be dominated by candidates with similar structures, stalling further~evolution. In such cases, instead of the mating operation, a mutation operation generates a new candidate as follows. First, a structure $G^{*}$ is selected from the population. Then, a new candidate structure $G^{**}$ is produced by randomly selecting a bond between two atoms within 1.8~\r{A} in $G^{*}$, and rotating it 90 degrees about an axis passing through the center of mass of $G^{*}$ and perpendicular to the bond. The new candidate structure $G^{**}$ is relaxed and replaces a candidate in the population if it has a lower energy. The choice between mutation or mating is based on the coordination number of atoms in the population: if all atoms have the same coordination number, the mutation is performed; otherwise, mating is chosen. A cutoff distance of 1.85 Å was used to calculate the coordination number.
The GA optimizations were performed for 10000 generations on different sized clusters C$_{N}$ ($4 \leq N \leq 64$, $N$ is~even) using MLBOP. The energies of the GA-optimized structures were recalculated with DFT for verification.

\begin{figure}[phtb!]
\centering
\includegraphics[clip,scale=0.37]{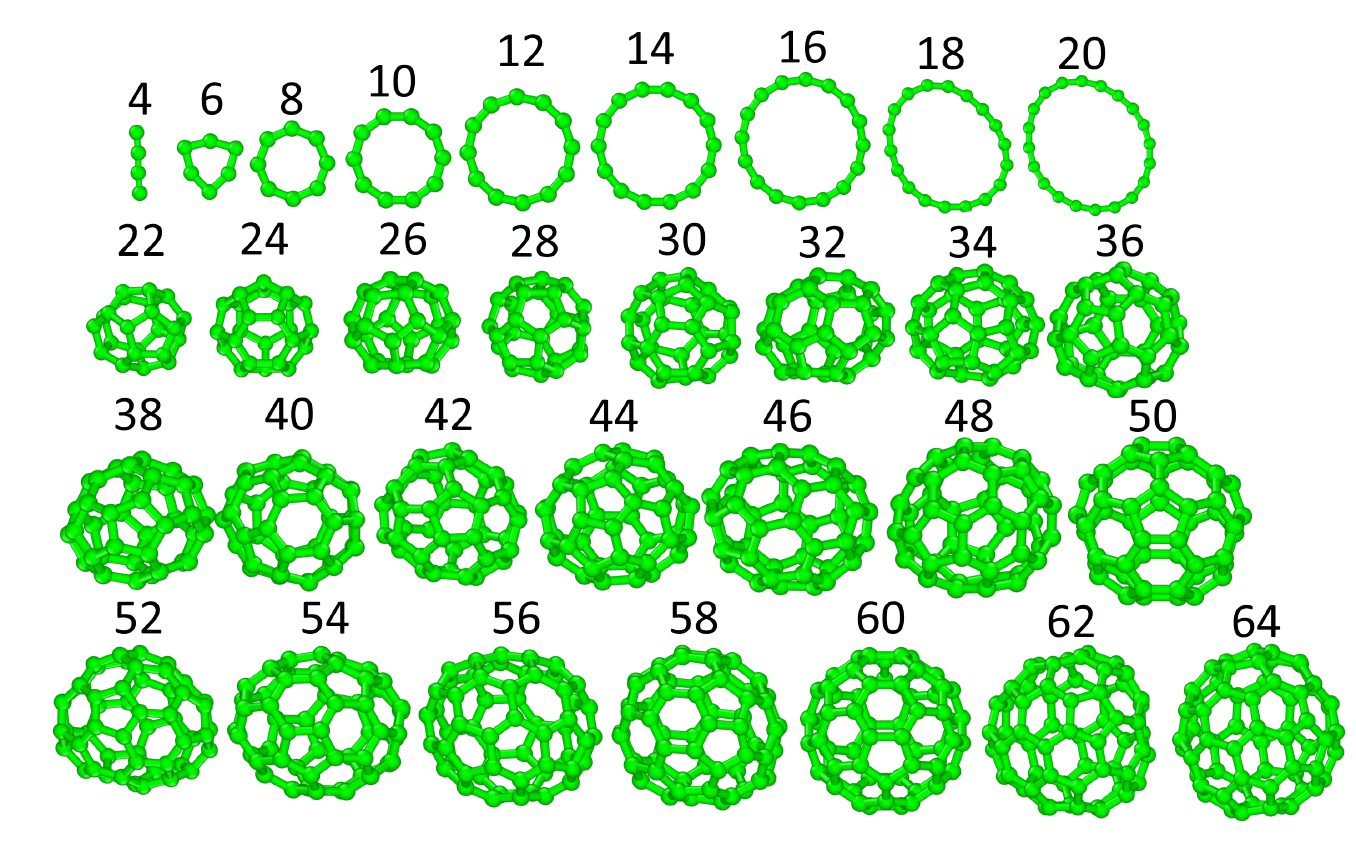}
\includegraphics[clip,scale=0.50]{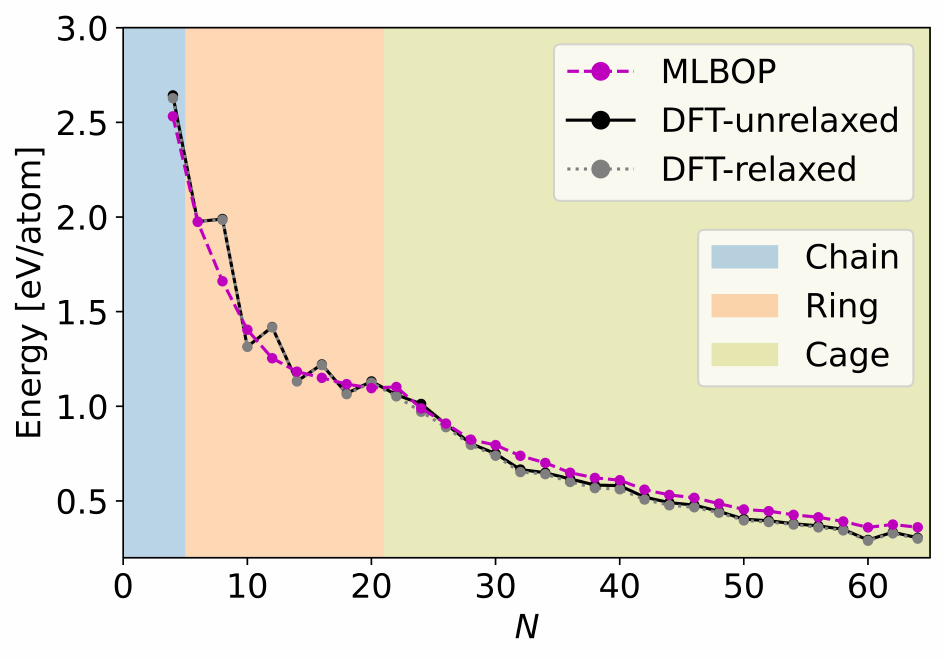}
\caption{
Top: GA-optimized structures of C$_{N}$ ($4 \leq N \leq 64$, $N$ is~even) obtained using MLBOP. Bottom: Energies (relative to monolayer graphene) of the GA-optimized structures obtained using MLBOP. The magenta and black points indicate the MLBOP and DFT (PBE-D3) energies of the GA-optimized structures, respectively. The gray points indicate the DFT energies of the DFT-relaxed structures starting from the GA-optimized structures. The blue, orange, and beige regions represent the size ranges in which C$_{N}$ is a chain, monocyclic ring, and cage, respectively.} 
\label{fig:gastructure}
\end{figure}

The top in Figure \ref{fig:gastructure} shows the GA-optimized structures obtained using MLBOP.
The structures optimized using MLBOP exhibit diverse forms depending on the cluster size: chain for $N \leq 4$, monocyclic ring for $6 \leq N \leq 20$, and cage for $22 \leq N \leq 64$, including the buckyball structure for C$_{60}$. This size-dependent structural~evolution from chains to rings, and~eventually to cages, is consistent with previous DFT studies \cite{10.1063/1.478414}. 
The bottom in Figure \ref{fig:gastructure} shows the energies of the GA-optimized structures calculated with MLBOP and DFT, and the DFT energies of the DFT-relaxed structures starting from the GA-optimized structures.
The DFT relaxation caused only small atomic displacements and energy differences in the GA-optimized structures obtained with MLBOP, indicating that MLBOP accurately predicts at least the local minima structures. The MLBOP energies of the cage structures are in good agreement with the DFT energies. 
In the DFT calculations, C$_{4N}$ rings are less stable compared to C$_{4N+2}$ rings due to differences in aromaticity. However, this instability is not reproduced by MLBOP; the MLBOP energy of C$_{4N}$ rings lies between that of C$_{4N-2}$ and C$_{4N+2}$. 
These indicate that the accuracy of MLBOP trained in this study is limited for small clusters with 20 atoms or less, suggesting room for improvement.

%For $22 \leq N \leq 60$, the GA-optimized structures obtained using the mod-Brenner potential and C-ACE are cages and C$_{60}$ is the buckyball as the MLBOP GA, but for $12 \leq N \leq 20$, graphitic sheets were predicted to be more stable than rings in contrast to the MLBOP GA results and DFT-recalculated energies, as shown in the insets in Figure \ref{fig:ga}, which is considered to be related to the underestimation of the graphene edge energy. For $N=6$, the ring structure obtained with MLBOP has D$_{3h}$ symmetry in agreement with the DFT ground state structure, while those obtained with the mod-Brenner potential and ACE have D$_{6h}$ symmetry. For $N=4$, the DFT calculation predicts the linear chain as the ground state structure. MLBOP can reproduce the DFT ground state structure of C$_{4}$, while the mod-Brenner and C-ACE predict the curved chain and square as the stable structures.

\subsection{Phase diagram}

\begin{figure}[htb!]
\centering
\includegraphics[clip,scale=0.52]{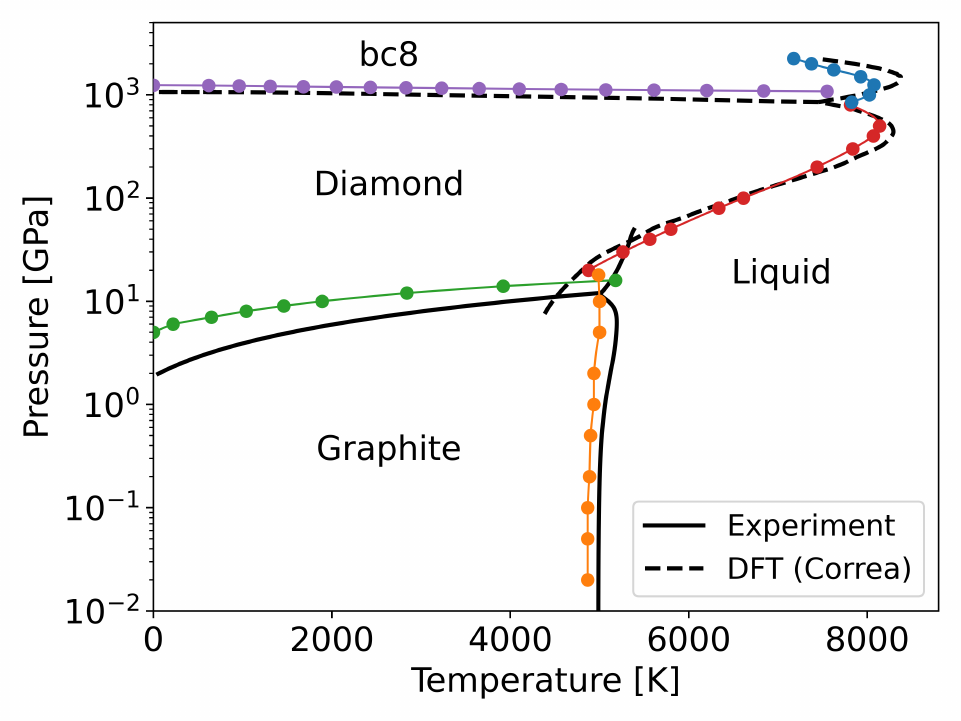}
\includegraphics[clip,scale=0.52]{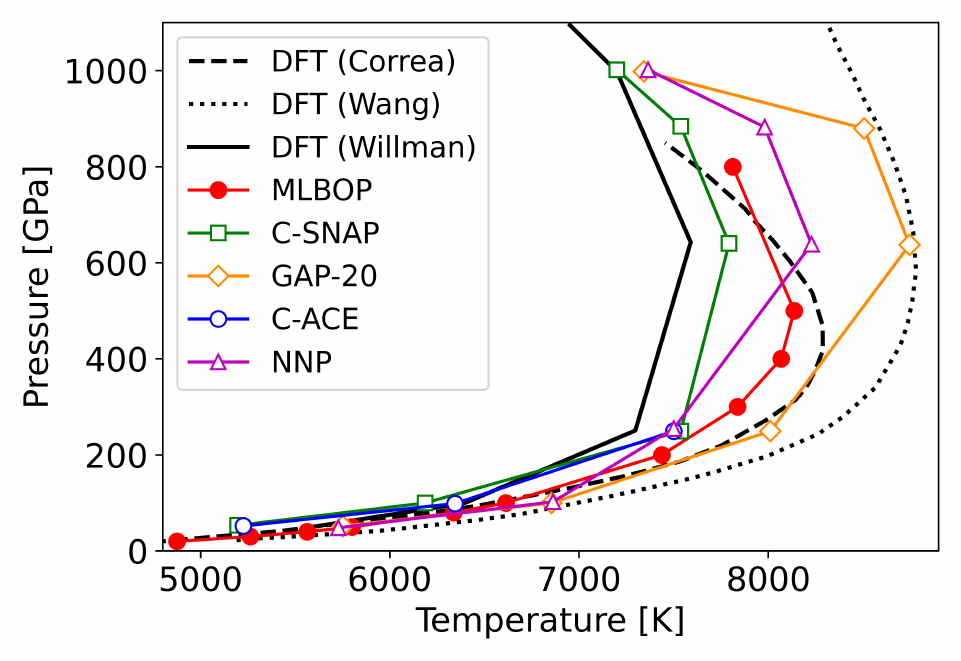}
\caption{Top: Phase diagram of carbon calculated with MLBOP. The green and purple dots indicate the graphite-diamond and the diamond-bc8 transition points, respectively. The orange, red, and blue dots indicate the melting points of graphite, diamond, and bc8, respectively. The colored lines are the transition and melting lines obtained as interpolations. The phase diagrams from the experiment (solid, black) \cite{BUNDY1996141} and the DFT calculation (dashed, black) by Correa et al. \cite{doi:10.1073/pnas.0510489103} are shown for comparison. Bottom: Magnified view of the diamond melting line. The diamond melting lines calculated by Wang et al. \cite{PhysRevLett.95.185701} using DFT, and by Willman et al. \cite{10.1063/5.0218705} using DFT, GAP-20, C-ACE, C-SNAP, and a Behler–Parrinello type NNP \cite{Cheng2023}, are also included for comparison.}
\label{fig:phase}
\end{figure}

To evaluate the performance of MLBOP in capturing phase behavior, we calculated the phase diagram of carbon using MLBOP at pressures from 0.02 GPa to 2500 GPa. To determine the melting points of graphite, diamond, and bc8, solid-liquid two-phase simulations \cite{PhysRevB.49.3109} were conducted. These simulations included 1008, 1024, and 864 atoms for graphite, diamond, and bc8, respectively, within orthogonal cells. The graphite-liquid interface was oriented perpendicular to the graphite layers, while the diamond-liquid and bc8-liquid interfaces were separated by the (100) surface. The two-phase simulations were performed using isothermal-isobaric (NPT) MD simulation for 40 ps, with a time step of 1 fs and varying temperatures in 25 K increments at each pressure. Melting points were determined as the solid-liquid equilibrium temperature. The solid-liquid phase transition was identified by monitoring the time history of the averaged Steinhardt order parameter \cite{PhysRevB.28.784}: specifically, Q$_3$ for the graphite-liquid system and Q$_6$ for the diamond- and bc8-liquid systems.
To obtain the graphite-diamond and diamond-bc8 transition lines, we calculated the Gibbs free energy using phonon calculations within the quasiharmonic approximation. The transition points were determined as temperatures at which the Gibbs free energies of the two structures are equal at each pressure. The phonon calculations were performed using phonopy \cite{phonopy-phono3py-JPCM,phonopy-phono3py-JPSJ}.
Top in Figure \ref{fig:phase} compares the phase diagram calculated with MLBOP with those from experiments \cite{BUNDY1996141} and DFT calculations \cite{doi:10.1073/pnas.0510489103}.
Overall, the MLBOP-calculated phase diagram shows good agreement with the DFT and experimental results.
The graphite-diamond transition line obtained with MLBOP is slightly shifted above the experimental transition line. The graphite-diamond transition point at 0 K calculated using MLBOP is 5.2 GPa, which is about 3 GPa higher than the experimental value. However, the MLBOP value is in agreement with the PBE-D3 value of 5.5 GPa, obtained using the same settings used to calculate the reference DFT energies. Thus, the shift in the graphite-diamond transition line is caused by inaccuracies in the DFT calculation method rather than fitting errors in MLBOP. Bottom in Figure \ref{fig:phase} shows the magnified view of the diamond melting line. The diamond melting lines calculated by Wang et al. \cite{PhysRevLett.95.185701} using DFT, and by Willman et al. \cite{10.1063/5.0218705} using DFT, C-SNAP, GAP-20, C-ACE, and a Behler–Parrinello type NNP \cite{Cheng2023}, are also shown for comparison. The melting line predicted by MLBOP has a maximum melting temperature of 8000 K, which is generally consistent with the range predicted by DFT calculations. Although the maximum melting points predicted by MLBOP, C-SNAP, GAP-20, and the NNP differ by up to approximately 1000 K, these models are considered to possess stable potential energy surfaces in the repulsive region, as demonstrated by their ability to sustain stable molecular dynamics simulations at high pressures.  While C-ACE shows good agreement with DFT calculations up to 300 GPa, its molecular dynamics simulations become unstable at higher pressures, as reported in Ref. \cite{10.1063/5.0218705}, indicating that its potential energy surface is not robust in the repulsive region.

%\subsection{Nucleation}

\subsection{Enthalpy-volume map of local minima structures}
\begin{figure*}[phtb!]
\centering
\includegraphics[clip,scale=0.20]{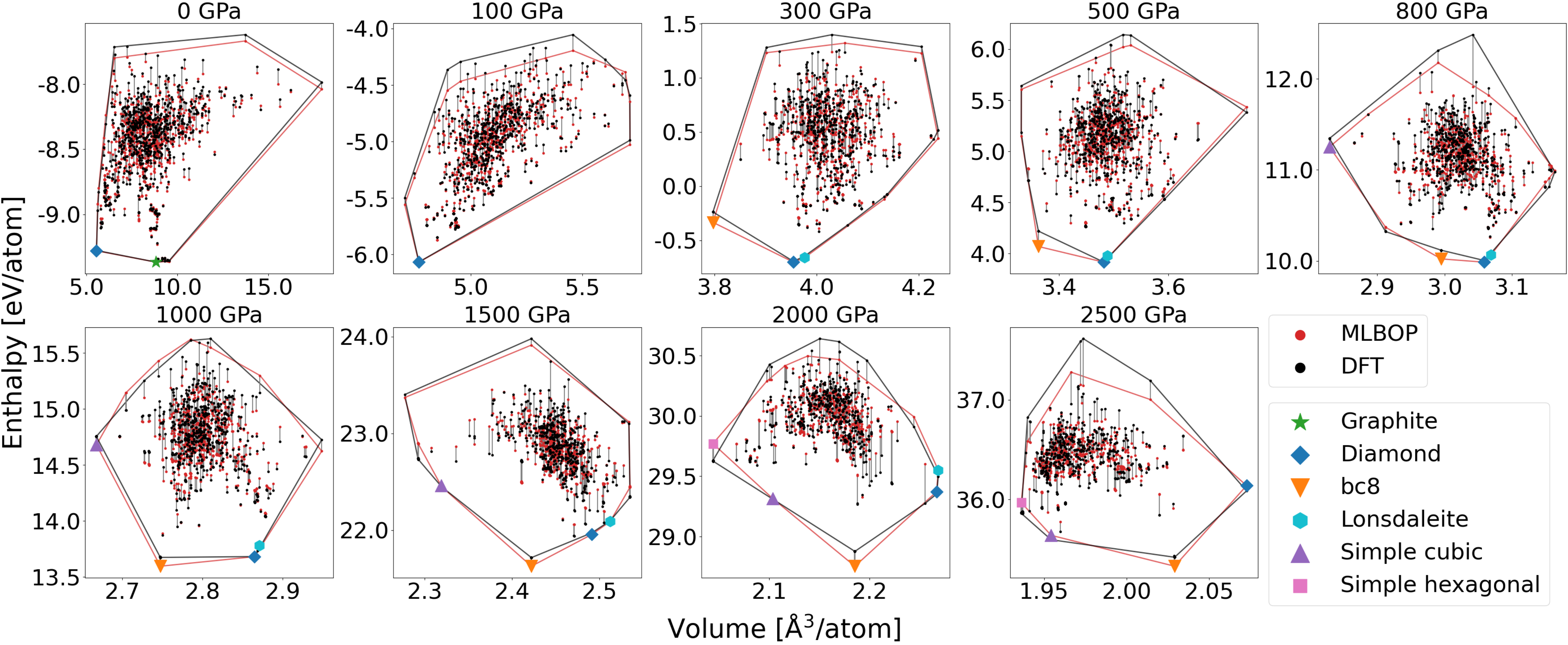}
\caption{Enthalpies and volumes of the MLBOP random structure search (RSS) minima structures at 0, 100, 300, 500, 800, 1000, 1500, 2000, and 2500 GPa. The red points indicate the MLBOP RSS minima and the black points indicate the DFT-recalculated enthalpies of the same minima structures. The red and black lines are the convex hulls surrounding the RSS minima points with the same colors. The convex hull points corresponding to AB-stacked graphite, diamond, bc8, Lonsdeleite, simple cubic, and simple hexagonal are represented by various markers.}
\label{fig:rss}
\end{figure*}

\begin{figure}[phtb!]
\centering
\includegraphics[clip,scale=0.38]{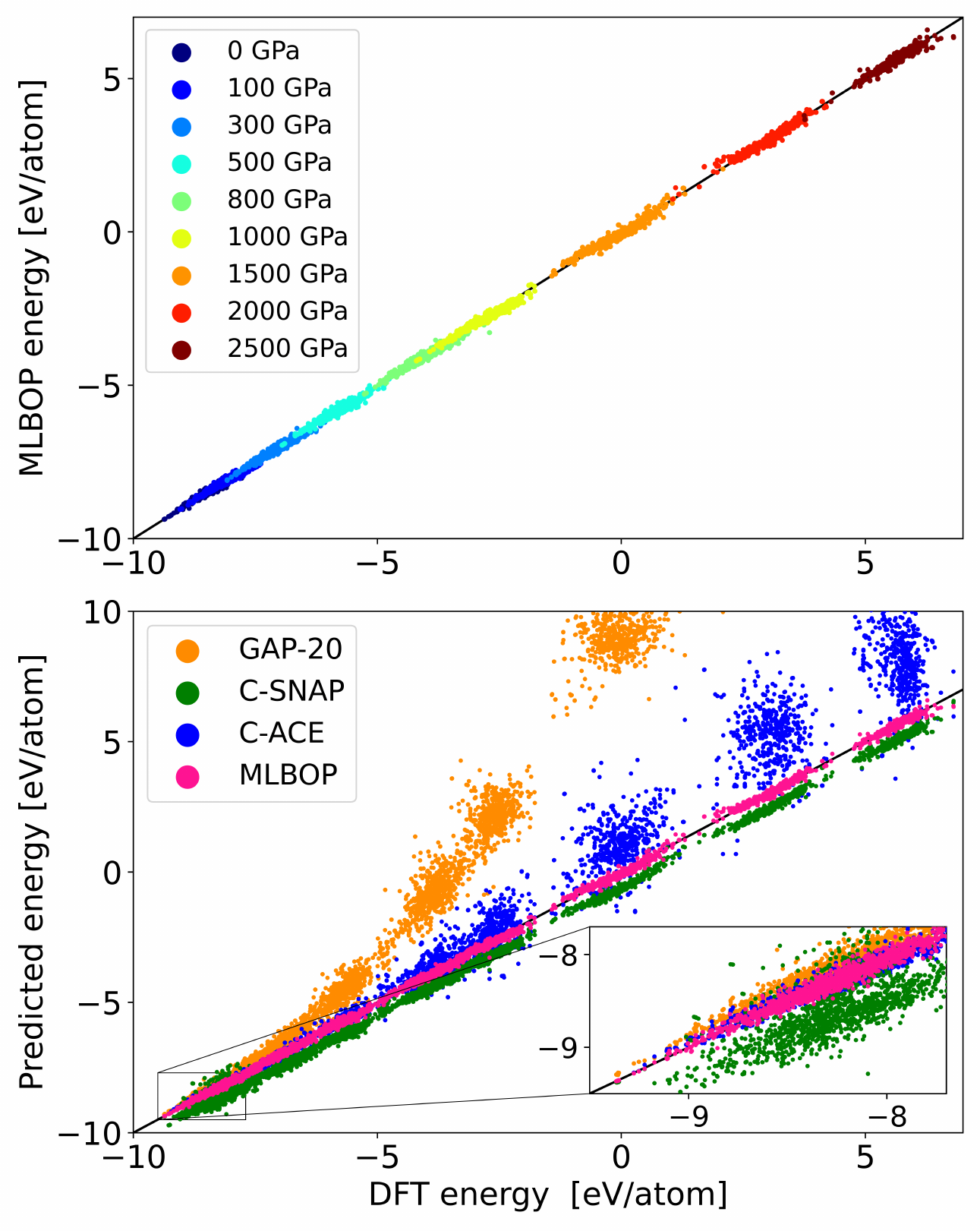}
\caption{Top: Comparison of the RSS minima energies between DFT and MLBOP. Bottom: Comparison of the energy prediction on the same RSS minima between GAP-20, C-SNAP, C-ACE, and MLBOP. All predicted values are shifted so that the graphene energy matches the value calculated using DFT. Inset shows the magnified view of low-energy region.}
\label{fig:rssplot}
\end{figure}

As shown in Figure \ref{fig:scadaene} and mentioned in Refs. \cite{10.1063/5.0091698,Marchant2023}, reported MLIPs for carbon often show some spurious ground state structures, which limits their applicability to structure searching.
To validate the correctness of the ground state structures predicted by MLBOP and its suitability for exploring the configuration space, we performed the RSS with applying external pressures of 0, 100, 300, 500, 800, 1000, 1500, 2000, and 2500 GPa. For each pressure, a total of 800 structures, 400 with 8 atoms and 400 with 16 atoms in the unit cell, were used as initial configurations for the RSS. These structures were generated by randomly placing atoms in a 3$\times$3$\times$3~\r{A}$^3$ cell with a minimum interatomic spacing of 1.1~\r{A}.
The atomic configurations and cell geometries were optimized using MLBOP and the FIRE algorithm \cite{PhysRevLett.97.170201} implemented in LAMMPS. During the optimization, the pressures in the normal (x, y, z) and shear (xz, xy, yz) directions were independently controlled, leading to a triclinic cell shape of the optimized structures.
The energies of the MLBOP RSS minima structures were recomputed with DFT using the PBE-D3 functional.
Figure \ref{fig:rss} shows the enthalpies and volumes of the RSS minima calculated with MLBOP. The convex hull shapes surrounding these minima are consistent with those from DFT calculations, indicating that MLBOP accurately reproduces the relative stability of structures at the convex hull points. The ground-state structures are AB-stacked graphite at 0 GPa, diamond between 100–800 GPa, and bc8 at higher pressures in agreement with DFT results. 
The top panel in Figure \ref{fig:rssplot} compares the energies of the MLBOP RSS minima structures between DFT (PBE-D3) and MLBOP. The bottom panel in Figure \ref{fig:rssplot} compares the energy prediction on the same RSS minima structures between GAP-20, C-ACE, C-SNAP, and MLBOP. Compared to other MLIPs, MLBOP accurately reproduced the DFT energies of the RSS minima structures over a wide range of pressures and achieved an accuracy of MAE 64.9~meV/atom. 

We also performed the RSS using C-ACE and GAP-20 starting from the same initial structures employed in the MLBOP RSS. For C-ACE, the majority of RSS structures collapsed to extremely small volumes at pressures exceeding 300 GPa with energies significantly lower than that of graphite due to the presence of spurious local minima in dense configurations. This unphysical energy drop in the repulsive extrapolation domains is a typical behavior of the ACE model and may be modified by replacing it with a pairwise core repulsion term as described in Ref. \cite{bochkarev2022efficient}. GAP-20 produced spurious ground-state phases such as strained diamond and strained hexagonal structures at pressures above 300 GPa, consistent with the findings reported in Ref. \cite{Marchant2023}. 
MLBOP did not produce spurious minima structures and correctly predicted ground state phases at each pressure, demonstrating the quality of the potential energy surface and the accuracy of the structural stability in the potential energy surface over a wide pressure range.

\section{Conclusion}
\label{sec:conclusion}
In this work, we constructed an MLIP model by combining a machine-learned bond-order with a Morse-type two-body energy function. Based on this model, we developed a general-purpose MLIP for carbon that provides a robust description of the potential energy surface and accurately reproduces the physical properties of various atomic structures. We demonstrated the performance of MLBOP in exploring the configuration space by carrying out the global structure search for clusters, calculating the phase diagram, and generating the enthalpy-volume maps of the RSS minima structures. These results show that it is possible to construct an MLIP that comprehensively covers the configuration space with a relatively small number of parameters within the formulation of the bond-order potential. 
We expect that MLBOP developed in this study will contribute to the discovery of new carbon materials with unique properties.

%At the same time, we also have several open problems to be addressed, as listed below:

\section*{Data availability}
The training and testing datasets listed in Table \ref{table:traindataset} and \ref{table:testdataset} are provided in Ref. \cite{data_large}. The PyTorch implementation of MLBOP is provided in Ref. \cite{data_large}. The input scripts, log files, and source codes of MLBOP, DeePMD, ACE, and SchNet for reproducing the training results are provided in Ref. \cite{data_large}. 
The LAMMPS implementation and parameters of the bonding energy part $E_\mathrm{bond}$ and the dispersion energy correction part $E_\mathrm{disp}$ of MLBOP are available in Ref. \cite{mlbop_github} and Ref. \cite{dftd3_github}, respectively. Other data and codes supporting the findings of this study are available from the contact author upon reasonable request.

\section*{Acknowledgements}
We acknowledge Center for Computational Materials Science, Institute for Materials Research, Tohoku University for the facilities of MASAMUNE-IMR. Additional computing facilities were provided by Initiative on Recommendation Program for Young Researchers and Woman Researchers, Information Technology Center, The University of Tokyo. A part of this work was supported by JSPS (KAKENHI JP21KK0087, JP22H01411, JP23H00174, JP23H05443) and JST (CREST JPMJCR20B5, SPRING JPMJSP2108).

\newpage
\bibliography{apssamp}% Produces the bibliography via BibTeX.

\end{document}